\documentclass[showpacs,preprintnumbers,amsmath,amssymb,twocolumn]{revtex4}

\usepackage{graphicx}
\usepackage{dcolumn}
\usepackage{txfonts,bm}
\usepackage{hyperref}
\usepackage{color}
\usepackage{epsfig}
\usepackage{multirow}
\usepackage{subfigure}

\begin{document}

\title{The Spin-orbit force, recoil corrections and possible
$B \bar{B}^{*}$ and $D \bar{D}^{*}$ molecular states}

\author{Lu Zhao$^{1}${\footnote{Email: Luzhao@pku.edu.cn}},
Li Ma$^{1}${\footnote{Email: lima@pku.edu.cn}}, Shi-Lin
Zhu$^{1,2}${\footnote{Email: zhusl@pku.edu.cn}}}

\affiliation{$^{1}$ Department of Physics and State Key Laboratory
of Nuclear Physics and Technology, Peking University, Beijing
100871, China\\
$^2$Collaborative Innovation Center of Quantum Matter, Beijing
100871, China }

\begin{abstract}

In the framework of the one boson exchange model, we have calculated
the effective potentials between two heavy mesons $B \bar{B}^{*}$
and $D \bar{D}^{*}$ from the t- and u-channel $\pi$, $\eta$, $\rho$,
$\omega$ and $\sigma$ meson exchange with four kinds of quantum
number: $I=0$, $J^{PC}=1^{++}$; $I=0$, $J^{PC}=1^{+-}$; $I=1$,
$J^{PC}=1^{++}$; $I=1$, $J^{PC}=1^{+-}$. We keep the recoil
corrections to the $B \bar{B}^{*}$ and $D \bar{D}^{*}$ system up to
$O(\frac{1}{M^2})$. The spin orbit force appears at
$O(\frac{1}{M})$, which turns out to be important for the very
loosely bound molecular states. Our numerical results show that the
momentum-related corrections are unfavorable to the formation of the
molecular states in the $I=0$, $J^{PC}=1^{++}$ and $I=1$,
$J^{PC}=1^{+-}$ channels in the $D \bar{D}^{*}$ systems.

\end{abstract}

 \maketitle{}

\section{INTRODUCTION}\label{introduction}

The charmonium spectroscopy has been studied extensively during the
last few years. The states below the open charm threshold are all
observed now while many states above the open charm threshold are
still missing. On the other hand, a large number of charmonium-like
states (or so called XYZ states) have been observed by experimental
collaborations such as Belle, Barbar, CDF, D0, LHCb, BESIII, CLEOc.
These XYZ states decay into the conventional charmonium, but some of
them do not fit into the quark model charmonium spectrum easily.
Especially the charged charmonium-like signals are the good
candidates of the exotic states, such as $Z(4430)$ observed in the
$\psi' \pi^{\pm}$ modes, $Z_1(4050)$, $Z_2(4250)$ in the $\chi_{c1}
\pi^{\pm}$ modes in the B meson decays
\cite{R.Mizuk:2008,Choi:2008,K.Chilikin:2013}, $Z_c(4025)^{\pm}$ in
the $\pi^{\pm}$ recoil mass spectrum and $Z_c(4020)^{\pm}$ in the
$\pi^{\pm}h_c$ mass spectrum \cite{Ablikim:2013}. Recently, the BES
Collaboration announced a charged structure $Z_c(3900)$ in the
$\pi^{\pm}J/\psi$ invariant mass spectrum of the process $e^+ e^-
\rightarrow \pi^+ \pi^- J/\psi$ at a center-of-mass (CM) energy of
$\sqrt{S}=4.260\pm0.001$GeV\cite{Ablikim:2013}. How to explain the
underlying structure of these charmonium-like states becomes an
important issue.

Many theoretical schemes were proposed to explain these XYZ states,
including the molecular states \cite{F.E.Close:2004,
M.B.Voloshin:2004, C.Y.Wong:2004, E.S.Swanson:2004,
N.A.Tornqvist:2004, Y.R.Liu:2010}, hybrid charmonium
\cite{B.A.Li:2005}, tetraquark states \cite{H.Hogaasen:2006,
D.Ebert:2006, N.Barnea:2006, Y.Cui:2007, R.D.Matheus:2007,
T.W.Chiu:2007}, dynamically generated resonances
\cite{D.Gamermann:2007}. Among the above schemes, the molecular
picture provides a plausible explanation since some XYZ states are
very close to the thresholds of a pair of charmed meson.

Since the first observation of $X(3872)$ by the Belle Collaboration
\cite{Choi:2003} in the exclusive decay process $B^{\pm}\!
\rightarrow\! K^{\pm}\pi^{+}\pi^{-} J/\psi$, its interpretation as a
molecular candidate of the $D \bar{D}^{*}$ system has been
investigated by many theoretical groups \cite{C.Y.Wong:2004,
E.S.Swanson:2004}\cite{M.T.AlFiky:2006, S.Fleming:2007,
E.Braaten:2007, C.Hanhart:2007, M.B.Voloshin:2007, P.Colangelo:2007,
M.Suzuki:2005, S.L.Zhu:2008}. Due to the same intriguing
near-threshold nature, the recently observed two charged
bottomonium-like states $Z_{b}(10610)$ and $Z_{b}(10650)$ by the
Belle observation \cite{I.Adachi:2011} were also interpreted as good
candidates of the $B \bar{B}^{*}$ and $B^* \bar{B}^{*}$ molecular
states \cite{Y.R.Liu:2008, X.Liu:2009, A.E.Bondar:2011,
D.Y.Chen:2011, Z.F.Sun:2011}. The newly observed $Z_c(3900)$ by
BESIII collaborations \cite{Ablikim:2013}, CLEOc \cite{T.Xiao:2013}
and Belle with ISR \cite{Z.Q.Liu:2013} is also close to the
threshold of $D \bar{D}^*$. In many references, it was interpreted
as the isovector partner of the well established isoscalar state
$X(3872)$ with the same quantum number $J^{P}=
1^{+}$.\cite{Q.Wang:2013, Z.G.Wang:2013, F.Aceti:2013}.

When investigating the possibility of $X(3872)$ as the $D
\bar{D}^{*}$ molecular state with $J^{PC}= 1^{++}$, the
one-pion-exchange (OPE) model and one-boson-exchange (OBE) model
were used to calculate the binding energy of the $D \bar{D}^{*}$
system in the Ref.\cite{N.Li:2012}. In Ref.\cite{Z.F.Sun:2011}, the
OBE model was applied to investigate the possibility of
$Z_{b}(10610)$ and $Z_{b}(10650)$ as the molecular states of the $B
\bar{B}^{*}$ and $B^* \bar{B}^{*}$ system.

With the exchange of the light pseudoscalar, vector and scalar
mesons, the OBE model provides an effective framework to describe
the interaction between two hadrons at different range. In the
previous work, the heavy quark symmetry is always invoked to
simplify the calculation in the derivation of the interaction
potential between two heavy mesons such as $D \bar{D}^{*}$ or $B
\bar{B}^{*}$. Moreover, the three momentum of the external particles
is sometimes ignored. Hence the resulting potential between the
heavy mesons depends on the exchanged momentum only. All the recoils
corrections were omitted.

The possible $D \bar{D}^{*}$ or $B \bar{B}^{*}$ molecular system is
very close to the two heavy meson threshold. The binding energy is
sometime quite small. Especially in the case of X(3872), its binding
energy may be less than 1 MeV if it turns out to be a $D
\bar{D}^{*}$ molecule. Compared to the tiny binding energy, the
higher order recoil corrections may turn out to be non-negligible.

In the present work, we keep the momentum of the initial and final
states explicitly and derive the effective potential using the
relativistic Lagrangian. We will keep the recoil corrections up to
the order $\frac{1}{M^2}$, where $M$ is the mass of the component in
the system. Especially the spin-orbit force first appears at ${O
(\frac{1}{M})}$. With the effective potentials with the explicit
recoil corrections ${O (\frac{1}{M^2})}$, we carefully investigate
the $D \bar{D}^{*}$ system with $I=0$, $J^{PC}= 1^{++}$ to measure
the $\frac{1}{M^2}$ correction for $X(3872)$, $D \bar{D}^{*}$ system
with $I=1$, $J^{PC}= 1^{+}$ for $Z_c(3900)$, and $B \bar{B}^{*}$
with $I=1$, $J^{PC}= 1^{+}$ for $Z_{b}(10610)$. Numerically, these
recoil corrections are quite important in the loosely bound heavy
meson systems. Especially, the recoil correction is comparable to
the binding energy in the case of X(3872).

This paper is organized as follows. We first introduce the formalism
of the derivation of the effective potential in Section
\ref{potential}. We present our numerical results in Section
\ref{Numerical}. The last section is the summary and discussion. We
collect some lengthy formulae in the appendix.

\section{The effective potential}\label{potential}
\begin{center}
\textbf{A. Wave function, Effective Lagrangian and Coupling
constants}
\end{center}

First, we construct the flavor wave functions of the isovector and
isoscalar molecular states composed of the $B\bar{B}^{*}$ and
$D\bar{D}^{*}$ as in Refs. \cite{Y.R.Liu:2008,X.Liu:2009}. The
flavor wave function of the $B\bar{B}^{*}$ system reads
\begin{equation}
\left\{
  \begin{array}{ll}
    |1,1\rangle = \frac{1}{\sqrt{2}}(|B^{*+}\bar{B}^0\rangle +
c|B^{+}\bar{B}^{*0}\rangle),  \\
    |1,-1\rangle = \frac{1}{\sqrt{2}}(|B^{*-}B^0\rangle +
c|B^{-}B^{*0}\rangle),  \\
    |1,0\rangle = \frac{1}{2}[(|B^{*+}B^-\rangle
-|B^{*0}\bar{B}^0\rangle) +
c(|B^{+}B^{*-}\rangle-|B^{0}\bar{B}^{*0}\rangle)],
  \end{array}
\right.
\end{equation}

\begin{equation}
|0,0\rangle = \frac{1}{2}[(|B^{*+}B^-\rangle
+|B^{*0}\bar{B}^0\rangle) +
c(|B^{+}B^{*-}\rangle+|B^{0}\bar{B}^{*0}\rangle)]
\end{equation}
where $c=\pm$ corresponds to C-parity $C=\mp$ respectively. For the
$D\bar{D}^{*}$ system
\begin{equation}
\left\{
  \begin{array}{ll}
    |1,1\rangle = \frac{1}{\sqrt{2}}(|\bar{D}^{*0}D^{+}\rangle +
c|\bar{D}^{0}D^{*+}\rangle),  \\
    |1,-1\rangle = \frac{1}{\sqrt{2}}(|D^{*0}D^-\rangle +
c|D^{0}D^{*-}\rangle),  \\
    |1,0\rangle = \frac{1}{2}[(|\bar{D}^{*0}D^0\rangle
-|D^{*-}D^+\rangle) +
c(|\bar{D}^{0}D^{*0}\rangle-|D^{-}D^{*+}\rangle)],
  \end{array}
\right.
\end{equation}

\begin{equation}
|0,0\rangle = \frac{1}{2}[(|\bar{D}^{*0}D^0\rangle
+|D^{*-}D^{+}\rangle) +
c(|\bar{D}^{0}D^{*0}\rangle+|D^{-}D^{*+}\rangle)]
\end{equation}

Since the C-parity of $Z_b(10610)^0$ is odd, we will take the
coefficient $c=+$ for the $B\bar{B}^{*}$ system. While the C parity
of $X(3872)$ was even, the $I=0$ $D\bar{D}^{*}$ system will take the
coefficient $c=-$. Moreover, we will consider both two C-parity
option for the $I=1$ $D\bar{D}^{*}$ system.

\begin{figure}[ht]
  \begin{center}
    \rotatebox{0}{\includegraphics*[width=0.30\textwidth]{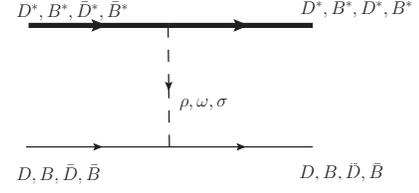}}
    \caption{ The direct-channel Feynman diagrams for both the $D\bar{D}^{*}$ and
$B\bar{B}^{*}$ systems at the tree level. The thick line represents
the vector state $D^*$, $B^*$, $\bar{D}^{*}$ or $\bar{B}^{*}$ while
the thin line stands for $D$, $B$, $\bar{D}$ and $\bar{B}$, .}
    \label{t-channel}
  \end{center}
\end{figure}

\begin{figure}[ht]
  \begin{center}
    \rotatebox{0}{\includegraphics*[width=0.35\textwidth]{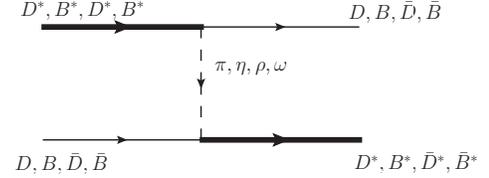}}
    \caption{ The cross-channel Feynman diagrams for both the $D\bar{D}^{*}$ and
$B\bar{B}^{*}$ systems at the tree level. Notations are the same as
in Fig. \ref{t-channel}}.
    \label{u-channel}
  \end{center}
\end{figure}

The meson exchange Feynman diagrams for both the $B\bar{B}^{*}$ and
$D\bar{D}^{*}$ systems at the tree level are shown in Fig.
\ref{t-channel} and Fig. \ref{u-channel}.

Based on the chiral symmetry, the Lagrangian for the pseudoscalar,
scalar and vector meson interaction with the heavy flavor mesons
reads
\begin{eqnarray}
\mathcal{L}_{P}&=&-i\frac{2g}{f_\pi}\bar{M}
P^{*\mu}_b\partial_{\mu}\phi_{ba}P^{\dag}_{a}+i\frac{2g}{f_\pi}\bar{M} P_b\partial_{\mu}\phi_{ba}P^{*\mu\dag}_{a} \nonumber\\
&-& \frac{g}{f_\pi}
P^{*\mu}_b\partial^{\alpha}\phi_{ba}\partial^{\beta}P^{*\nu\dag}_{a}\epsilon_{\mu\nu\alpha\beta}
+ \frac{g}{f_\pi}
\partial^{\beta}P^{*\mu}_b\partial^{\alpha}\phi_{ba}P^{*\nu\dag}_{a}\epsilon_{\mu\nu\alpha\beta},\label{pseudo-exchange}
\end{eqnarray}

\begin{eqnarray}
\widetilde{\mathcal{L}_{P}}&=&-i\frac{2g}{f_\pi}\bar{M}\widetilde{P^{\dag}_{a}}
\partial_{\mu}\phi_{ab}\widetilde{P^{*\mu}_b}-i\frac{2g}{f_\pi}\bar{M}\widetilde{P^{*\mu\dag}_{a}}\partial_{\mu}\phi_{ab}\widetilde{P_b} \nonumber\\
&+& \frac{g}{f_\pi}
\partial^{\beta}\widetilde{P^{*\mu\dag}_{a}}\partial^{\alpha}\phi_{ab}\widetilde{P^{*\nu}_b}\epsilon_{\mu\nu\alpha\beta}
-\frac{g}{f_\pi}
\widetilde{P^{*\mu\dag}_{a}}\partial^{\alpha}\phi_{ab}\partial^{\beta}\widetilde{P^{*\nu}_b}\epsilon_{\mu\nu\alpha\beta},\label{anti-pseudo-exchange}
\end{eqnarray}

\begin{eqnarray}
\mathcal{L}_{V}&=&i\frac{\beta g_v}{\sqrt{2}} P_b
V^{\mu}_{ba}\partial_{\mu}P^{\dag}_{a}-i\frac{\beta g_v}{\sqrt{2}}
\partial_{\mu}P_b V^{\mu}_{ba}P^{\dag}_{a} \nonumber\\
&-&i\sqrt{2}\lambda g_v\epsilon_{\mu\alpha\beta\nu}\partial^{\mu}P_b
\partial^{\alpha}V^{\beta}_{ba}P^{*\nu\dag}_{a}\nonumber\\
&-&i\sqrt{2}\lambda g_v \epsilon_{\mu\alpha\beta\nu}P^{*\mu}_b
\partial^{\alpha}V^{\beta}_{ba}\partial^{\nu}P^{\dag}_{a}\nonumber\\
&-&i\frac{\beta g_v}{\sqrt{2}} P^{*\nu}_b
V^{\mu}_{ba}\partial_{\mu}P^{*\dag}_{\nu a}+ i\frac{\beta
g_v}{\sqrt{2}} \partial_{\mu}P^{*\nu}_b V^{\mu}_{ba}P^{*\dag}_{\nu a}\nonumber\\
&-&i2\sqrt{2}\lambda g_v \bar{M^*}P^{*\mu}_b
(\partial_{\mu}V_{\nu}-\partial_{\nu}V_{\mu})_{ba}P^{*\nu\dag}_a,\label{vector-exchange}
\end{eqnarray}

\begin{eqnarray}
\widetilde{\mathcal{L}_{V}}&=&-i\frac{\beta g_v}{\sqrt{2}}
\partial_{\mu}\widetilde{P^{\dag}_{a}} V^{\mu}_{ab}\widetilde{P_b}+
i\frac{\beta g_v}{\sqrt{2}} \widetilde{P^{\dag}_{a}} V^{\mu}_{ab}\partial_{\mu}\widetilde{P_b} \nonumber\\
&+&i\sqrt{2}\lambda g_v\epsilon_{\mu\alpha\beta\nu}
\widetilde{P^{*\mu\dag}_{a}}\partial^{\alpha}V^{\beta}_{ab}\partial^{\nu}\widetilde{P_b}\nonumber\\
&+&i\sqrt{2}\lambda g_v \epsilon_{\mu\alpha\beta\nu}\partial^{\mu}
\widetilde{P^{\dag}_{a}}\partial^{\alpha}V^{\beta}_{ab}\widetilde{P^{*\nu}_b}\nonumber\\
&+&i\frac{\beta g_v}{\sqrt{2}}\partial_{\mu}
\widetilde{P^{*\dag}_{\nu a}} V^{\mu}_{ba} \widetilde{P^{*\nu}_b}-
i\frac{\beta g_v}{\sqrt{2}}\widetilde{P^{*\dag}_{\nu a}}
 V^{\mu}_{ab}\partial_{\mu} \widetilde{P^{*\nu}_b}\nonumber\\
&-&i2\sqrt{2}\lambda g_v \bar{M^*}\widetilde{P^{*\mu\dag}_a}
(\partial_{\mu}V_{\nu}-\partial_{\nu}V_{\mu})_{ab}\widetilde{P^{*\nu}_b},\label{anti-vector-exchange}
\end{eqnarray}

\begin{eqnarray}
\mathcal{L}_{S}=-2g_s \bar{M}P_b\sigma P^{\dag}_{b}+ 2g_s
\bar{M^*}P^{*\mu}_b\sigma P^{*\dag}_{\mu b}\label{scalar-exchange}
\end{eqnarray}

\begin{eqnarray}
\widetilde{\mathcal{L}_{S}}=-2g_s
\bar{M}\widetilde{P^{\dag}_a}\sigma \widetilde{P_{a}}+ 2g_s
\bar{M^*}\widetilde{P^{*\dag}_{\mu a}}\sigma
\widetilde{P^{*\mu}_a}\label{anti-scalar-exchange}
\end{eqnarray}
where the heavy flavor meson fields $P$ and $P^*$ represent $P=(D^0,
D^+)$ or $(B^-, \bar{B}^0)$ and $P^*=(D^{*0}, D^{*+})$ or $(B^{*-},
\bar{B}^{*0})$. Its corresponding heavy anti-meson fields
$\widetilde{P}$ and $\widetilde{P}^*$ represent
$\widetilde{P}=(\bar{D}^0,D^-)$ or $(B^+, B^0)$ and
$\widetilde{P}^*=(\bar{D}^{*0},D^{*-})$ or $(B^{*+}, B^{*0})$.
$\phi$, $V$ represent the the exchanged pseudoscalar and vector
meson matrices, $\sigma$ is the only scalar meson interacting with
the heavy flavor meson.

\begin{eqnarray}
\phi=\left(
         \begin{array}{cc}
           \frac{\pi^0}{\sqrt{2}}+\frac{\eta}{\sqrt{6}} & \pi^+ \\
           \pi^- & -\frac{\pi^0}{\sqrt{2}}+\frac{\eta}{\sqrt{6}} \\
         \end{array}
       \right)
\end{eqnarray}

\begin{eqnarray}
V=\left(
         \begin{array}{cc}
           \frac{\rho^0}{\sqrt{2}}+\frac{\omega}{\sqrt{2}} & \rho^+ \\
           \rho^- & -\frac{\rho^0}{\sqrt{2}}+\frac{\omega}{\sqrt{2}} \\
         \end{array}
       \right)
\end{eqnarray}

According the OBE model, five mesons ( $\pi$, $\sigma$, $\rho$,
$\omega$ and $\eta$) contribute to the effective potential. In the
$D\bar{D}^{*}$ and $B\bar{B}^{*}$ systems we considered, the
potentials are the same for the three isovector states in Eqs.
(1)$\sim$(4) with the exact isospin symmetry. Expanding the
Lagrangian densities in Eqs. (5)$\sim$(10) leads to each meson's
contribution for the two coupled channels. These channel-dependent
coefficients are listed in Table \ref{tab:channel-coeff}. The pionic
coupling constant $g\!=\!0.59$ is extracted from the width of
$D^{*+}$\cite{S.Ahmed:2001}. $f_{\pi}=132 MeV$ is the pion decay
constant. According the vector meson dominance mechanism, the
parameters $g_v$ and $\beta$ can be determined as $g_v=5.8$ and
$\beta=0.9$. At the same time, by matching the form factor obtained
from the light cone sum rule and that calculated from the lattice
QCD, we can get $\lambda=0.56 GeV^{-1}$\cite{C.Isola:2003,
M.Bando:1988}. The coupling constant related to the scalar meson
exchange is $g_s=g_{\pi}/2\sqrt{6}$ with $g_{\pi}=3.73$
\cite{X.Liu:2009, A.F.Falk:1992}. All these parameters are listed in
Table \ref{tab:coupling-constant}.

\begin{table}[htbp]
\caption{The coupling constants and masses of the heavy mesons and
the exchanged light mesons used in our calculation. The masses of
the mesons are taken from the PDG \cite{PDG}}
\label{tab:coupling-constant}
\begin{center}
\begin{tabular}{c | c  | c }
\hline \hline  & {mass(MeV)} & {coupling constants} \\
\cline{2-3} \hline
\multirow{2}{*}{pseudoscalar} & $m_{\pi}=134.98$  &  $g=0.59$  \\

                                      &  $m_{\eta}=547.85$ &  $f_{\pi}=132 MeV$  \\
\hline
\multirow{3}{*}{vector} & $m_{\rho}=775.49$       &  $g_v=5.8$  \\

                                & $m_{\omega}=782.65$     &  $\beta=0.9$  \\

                                &                         &  $\lambda=0.56 GeV^{-1}$  \\
\hline
\multirow{2}{*}{scalar} & $m_{\sigma}=600$    &  $g_s=g_{\pi}/2\sqrt{6}$  \\

                                &   &  $g_{\pi}=3.73$  \\
\hline\hline
\multirow{4}{*}{heavy flavor } & $m_{D}=1864.9$    &    \\

                                &  $m_{D^*}=2010.0$     &   \\
                                &  $m_{B}=5279.0$     &   \\
                                &  $m_{B^*}=5325$     &   \\
\hline\hline
\end{tabular}
\end{center}
\end{table}

\begin{table}[htbp]
\caption{coefficients } \label{tab:channel-coeff}
\begin{center}
\begin{tabular}{c| c  | c c c | c c c c }
\hline \hline & \multirow{2}{*}{isospin}   & \multicolumn{3}{c|}{direct-channel }& \multicolumn{4}{c}{cross-channel} \\
\cline{3-9}
& & $~~~\rho~~~$ & $~~~\omega~~~$ & $~~~\sigma~~~$ & $~~~\rho~~~$ & $~~~\omega~~~$ & $~~~\pi~~~$ & $~~~\eta~~~$\\
\hline
\multirow{2}{*}{$D\bar{D}^{*}$} & $I=1$ &   ~-1/2~ &  ~1/2~ & ~1~ & ~$-c/2$~  & ~$c/2$~ & $-c/2$  &  $c/6$  \\
\cline{2-9}
                                & $I=0$ &   ~3/2~  & ~1/2~  & ~1~ & ~$3c/2$~  & ~$c/2$~ & $3c/2$  &  $c/6$  \\
\hline
\multirow{2}{*}{$B\bar{B}^{*}$} & $I=1$ &   ~-1/2~ &  ~1/2~ & ~1~ & ~$-c/2$~  & ~$c/2$~ & $-c/2$  &  $c/6$  \\
\cline{2-9}
                                & $I=0$ &   ~3/2~  & ~1/2~  & ~1~  & ~$3c/2$~ &  ~$c/2$~ & $3c/2$  &  $c/6$  \\
\hline\hline
\end{tabular}
\end{center}
\end{table}

In order to include all the momentum-related terms in our
calculation, we need introduce the polarization vectors of the
vector mesons. The polarization vector at its rest frame is
\begin{eqnarray}\label{pol}
\epsilon_{\lambda}=(0,\vec{\epsilon_{\lambda}})
\end{eqnarray}
We need to make a lorentz boost to Eq. \ref{pol} to derive the
polarization vector in the laboratory frame
\begin{eqnarray}
\epsilon^{lab}_{\lambda}=(\frac{\vec{p}\cdot\vec{\epsilon_{\lambda}}}{m},
\vec{\epsilon_{\lambda}}+\frac{\vec{p}\cdot
(\vec{p}\cdot\vec{\epsilon_{\lambda}})}{m(P_0 + m)})
\end{eqnarray}
where $p=(p_0,\mathbf{p})$ is the particle's 4-momentum in the
laboratory frame and $m$ is the mass of the particle.

\begin{center}
\textbf{B. Effective potential }
\end{center}

Together with the wave function and Feynman diagram, we can derive
the relativistic scattering amplitude at the tree level
\begin{equation}
\langle f | S | i \rangle = \delta_{fi} + i \langle f | T | i
\rangle = \delta_{fi} + (2\pi)^4\delta^4(p_f-p_i) i M_{fi},
\end{equation}
where the T-matrix is the interaction part of the S-matrix and M is
defined as the invariant matrix element. After applying Bonn
approximation on the Lippmann-Schwinger equation, the S-matrix reads
\begin{equation}
\langle f | S | i \rangle = \delta_{fi} - 2\pi \delta(E_f-E_i) i
V_{fi}
\end{equation}
with $V_{fi}$ being the effective potential. Considering the
different normalization conventions used for the scattering
amplitude $M_{fi}$, $T$-matrix $T_{fi}$ and $V_{fi}$, we have
\begin{equation}
V_{fi}=-\frac{M_{fi}}{\sqrt{ \mathop\prod\limits_{f}2{p_f}^0
\mathop\prod\limits_{i} 2{p_i}^0}}\approx -\frac{M_{fi}}
{\sqrt{\mathop\prod\limits_{f} 2{m_f}^0 \mathop\prod\limits_{i}
2{m_i}^0}}
\end{equation}
where $p_{f(i)}$ denotes the four momentum of the final (initial)
state.

During our calculation, $P_1(E_1,\vec{p})$ and $P_2(E_2,-\vec{p})$
denote the four momenta of the initial particles in the center mass
system, while $P_3(E_3,\vec{p'})$ and $P_4(E_4,-\vec{p'})$ denote
the four momenta of the final particles, respectively.
\begin{equation}
q=P_3-P_1=(E_3-E_1,\vec{p'}-\vec{p})=(E_2-E_4,\vec{q})
\end{equation}
is the transferred four momentum or the four momentum of the meson
propagator. For convenience, we always use
\begin{equation}
\vec{q}=\vec{p'}-\vec{p}
\end{equation}
and
\begin{equation}
\vec{k}=\frac{1}{2}(\vec{p'}+\vec{p})
\end{equation}
instead of $\vec{p'}$ and $\vec{p}$ in the practical calculation.

In the OBE model, each vertex in the Feynman diagram needs a form
factor to suppress the high momentum contribution. We take the
conventional form for the form factor as in the Bonn potential
model.
\begin{equation}
F(q)=\frac{\Lambda^2-m_{\alpha}^2}{\Lambda^2-q^2}=\frac{\Lambda^2-m_{\alpha}^2}{{\tilde{\Lambda}}^2+\vec{q}^2}
\end{equation}
$m_\alpha$ is the mass of the exchanged meson and
\begin{equation}
\tilde{\Lambda}^2=\Lambda^2-(m^*-m)^2
\end{equation}
where $m$ and $m^*$ is the mass of the heavy flavor meson $D$ and
$D^*$ or $B$ and $B^*$. So far, the effective potential is in the
momentum space. In order to solve the time independent
Schr\"{o}dinger equation in the coordinate space, we need to make
the Fourier transformation to $V(\vec{q},\vec{k})$. The details of
the Fourier transformations are presented in the Appendix.

All the meson exchanged potentials for $B\bar{B}^{*}$ and
$D\bar{D}^{*}$ are the same, except the $\pi$ exchange potential.
The $\pi$ mass is larger than the mass difference of $B$ and
$\bar{B}^*$ but smaller than that of $D$ and $\bar{D}^*$.

The expressions of the direct-channel effective potential through
exchanging the $\sigma$, $\rho$ mesons are
\begin{eqnarray}
V_{\sigma}&=&-C_{\sigma}g^2_s(\vec{\epsilon_b}\cdot
\vec{\epsilon_a}^{\dag})F_{1t\sigma}\nonumber\\
&~&-C_{\sigma}g^2_s \frac{1}{2m^{*2}}(F_{3t1\sigma}+F_{3t2\sigma})\nonumber\\
&~&+C_{\sigma}g^2_s \frac{1}{2m^{*2}}\frac{(\vec{\epsilon_b}\times
\vec{\epsilon_a}^{\dag})\cdot \vec{L}}{i}F_{5t\sigma}
\end{eqnarray}

\begin{eqnarray}
V_{\rho}&=&-C_{\rho}\beta^2 g^2_v \frac{\vec{\epsilon_b}\cdot
\vec{\epsilon_a}^{\dag}}{2}F_{1t\rho}\nonumber\\
&~&+C_{\rho}(\frac{\lambda \beta g^2_v}{m^*}-\frac{\beta^2 g^2_v}{4m^{*2}} )(F_{3t1\rho}+F_{3t2\rho})\nonumber\\
&~&-C_{\rho}\beta^2 g^2_v \frac{\vec{\epsilon_b}\cdot
\vec{\epsilon_a}^{\dag}}{2m m^*}[F_{4t1\rho}+\{-\frac{1}{2}\nabla^2, F_{4t2\rho}\}]\nonumber\\
&~&+C_{\rho}( \frac{\beta^2 g^2_v }{4m^{*2}}-\lambda \beta
g^2_v\frac{m^* + m}{m m^*})\frac{(\vec{\epsilon_b}\times
\vec{\epsilon_a}^{\dag})\cdot \vec{L}}{i}F_{5t\rho}
\end{eqnarray}
The $\omega$ and $\rho$ meson exchange potentials have the same form
except that the meson mass and channel-dependent coefficients are
different.

The expression of the cross-channel effective potential through
exchanging the $\pi$ meson in the $B\bar{B}^{*}$ system is
\begin{eqnarray}
V_{\pi}&=&C C_{\pi} \frac{g^2_{\pi}}{f^2_{\pi}}\frac{(m^* + m)^2}{4 m^{*2}}(F_{3u1\pi}+F_{3u2\pi})\nonumber\\
&~&+C C_{\pi} \frac{g^2_{\pi}}{f^2_{\pi}} \frac{m^{*2}-m^2}{2m^{*2}}
\frac{(\vec{\epsilon_b}\times \vec{\epsilon_a}^{\dag})\cdot
\vec{L}}{i}F_{5u\pi}\nonumber\\
&~&-C C_{\pi}\frac{g^2_{\pi}}{f^2_{\pi}}\frac{(m^* -
m)^2}{m^{*2}}(F_{6u1}+F_{6u2\pi}\nabla + F_{6u3\pi}\nabla^2)
\end{eqnarray}

The expression of the cross-channel effective potential through
exchanging the $\pi$ meson in the $D\bar{D}^{*}$ system is
\begin{eqnarray}
V'_{\pi}&=&C C_{\pi} \frac{g^2_{\pi}}{f^2_{\pi}}\frac{(m^* + m)^2}{4 m^{*2}}(F'_{3u1\pi}+F'_{3u2\pi})\nonumber\\
&~&+C C_{\pi} \frac{g^2_{\pi}}{f^2_{\pi}} \frac{m^{*2}-m^2}{2m^{*2}}
\frac{(\vec{\epsilon_b}\times \vec{\epsilon_a}^{\dag})\cdot
\vec{L}}{i}F'_{5u\pi}\nonumber\\
&~&-C C_{\pi}\frac{g^2_{\pi}}{f^2_{\pi}}\frac{(m^* -
m)^2}{m^{*2}}(F'_{6u1}+F'_{6u2\pi}\nabla + F'_{6u3\pi}\nabla^2)
\end{eqnarray}

The expression of the cross-channel effective potential through
exchanging the $\rho$ meson is
\begin{eqnarray}
V_{\rho}&=&-C C_{\rho}\lambda^2 g^2_v \frac{(m^* + m)^2}{2m m^*}
(\vec{\epsilon_b}\cdot \vec{\epsilon_a}^{\dag})F_{2u\rho}\nonumber\\
&~&+C C_{\rho}\lambda^2 g^2_v \frac{(2m^*-m)(m^* + m)^2}{2m^{*3}}(F_{3u1\rho}+F_{3u2\rho})\nonumber\\
&~&+C C_{\rho}\lambda^2 g^2_v \frac{2(m^*-m)^2}{m m^*}
\vec{\epsilon_b}\cdot\vec{\epsilon_a}^{\dag} [F_{4u1\rho}+\{-\frac{1}{2}\nabla^2, F_{4u2\rho}\}]\nonumber\\
&~&-C C_{\rho}\lambda^2 g^2_v \frac{m(m^{*2} - m^2)}{m^{*3}}
\frac{(\vec{\epsilon_b}\times \vec{\epsilon_a}^{\dag})\cdot
\vec{L}}{i}F_{5u\rho}\nonumber\\
&~&+C C_{\rho}\lambda^2 g^2_v \frac{2(2m^*+m)(m^* -
m)^2}{m^{*3}}(F_{6u1\rho}+ F_{6u2\rho}\nabla\nonumber\\
&~&+F_{6u3\rho}\nabla^2)
\end{eqnarray}
Similarly, the $\eta$ and $\pi$ meson exchange potential has the
same form in the $B {\bar B}^{*}$ system. The potential from the
$\omega$ and $\rho$ meson exchange is also similar except the meson
mass and channel-dependent coefficients. The explicit forms of
$\mathcal{F}_{\mu t \alpha}$,$\mathcal{F}_{\mu u
 \alpha}$,$\mathcal{F}_{\mu t \nu \alpha}$, $\mathcal{F}_{\mu u
\nu \alpha}$, $\mathcal{F'}_{\mu u \nu \alpha}$ are shown in the
Appendix.

In our calculation, we explicitly consider the external momentum of
the initial and final states. Due to the recoil corrections, several
new terms appear which were omitted in the heavy quark symmetry
limit. These momentum dependent terms are related to the momentum
$\vec{k}=\frac{1}{2}(\vec{p'}+\vec{p})$:
\begin{eqnarray}
\frac{\vec{k}^2}{\vec{q}^2+m_{\alpha}^2}
\end{eqnarray}
and
\begin{eqnarray}\label{so}
{\frac{i\vec{S}\cdot\vec{k}\times\vec{q}}{\vec{q}^2+m_{\alpha}^2}}
\end{eqnarray}
and
\begin{eqnarray}\label{t2}
\frac{(\vec{\epsilon_b}\cdot\vec{k})(\vec{\epsilon_a}^{\dag}
\cdot\vec{k})}{\vec{p}^2+m_{\alpha}^2}
\end{eqnarray}
where $\vec{S}=-i(\vec{\epsilon_b}\times\vec{\epsilon_a}^{\dag})$.
The term in Eq. (\ref{so}) is the well-known spin orbit force. The
term in Eq. (\ref{t2}) depends on the spin and results in the
momentum-related operator $\nabla$, $\nabla^2$. The Fourier
transformation of the above new interaction terms are also shown in
the Appendix. In short, all the terms in the effective potentials in
the form of $F_{4t1\rho}$, $F'_{5u\pi}$, $(F_{6u1\rho}+
F_{6u2\rho}\nabla +F_{6u3\rho}\nabla^2)$ etc with the sub-indices
$4,5,6$ arise from the recoil corrections and vanish when the heavy
meson mass $m, m^*$ goes to infinity. Especially, the spin orbit
force appears at $O(1/M)$!

\begin{center}
\textbf{C.  Schr\"{o}dinger equation }
\end{center}
With the effective potential $V(\vec{r})$ in Eqs. (23) $\sim$ (27),
we are able to study the binding property of the system by solving
the Schr\"{o}dinger Equation
\begin{equation}
(-\frac{\hbar^2}{2\mu}\nabla^{2}+V(\vec{r})-E)\Psi(\vec{r})=0,
\label{eq:schrod}
\end{equation}
where $\Psi(\vec{r})$ is the total wave function of the system. The
total spin of the system $S=1$ and the orbital angular momenta $L=0$
and $L=2$. Thus the wave function $\Psi(\vec{r})$ should have the
following form
\begin{equation}
\Psi(\vec{r})=\psi_S(\vec{r})+\psi_D(\vec{r}),
\end{equation}
where $\psi_S(\vec{r})$ and $\psi_D(\vec{r})$ are the $S$-wave and
$D$-wave functions, respectively. In the matrix method, we use
Laguerre polynomials as a set of orthogonal basis
\begin{equation}
\chi_{nl}(r)=\sqrt{\frac{(2\lambda)^(2l+3) n!}{\Gamma(2l+3+n)}}r^l
e^{-\lambda r}L^{2l+2}_n (2\lambda r), n=1,2,3...
\end{equation}
with a normalization condition of
\begin{equation}
\int^\infty _0 \chi_{im}(r) \chi_{in}(r) r^2
dr=\delta_{ij}\delta_{mn}.
\end{equation}
We expand the total wave function as
\begin{equation}
\Psi(\vec{r})=\sum^{n-1}_{i=0}a_i \chi_{i 0}(r)\phi_S +
\sum^{n-1}_{p=0}b_p \chi_{p 2}(r)\phi_D, \label{eq:tottrail}
\end{equation}
where $\phi_S$ and $\phi_D$ are the angular part of the spin and
orbital wave function for the $S$- and $D$-states, respectively.
$a_i$ and $b_i$ are the corresponding expansion coefficients.

In the practical calculation, we detach the terms related to the
kinetic-energy-operator $\nabla^{2}$ and $\nabla$ from $V(\vec{r})$
and re-write Eq. (\ref{eq:schrod}) as
\begin{eqnarray}
&~&(-\frac{\hbar^2}{2\mu}\nabla^{2}-\frac{\hbar^2}{2\mu}[\nabla^2
\alpha(r)+\alpha(r)\nabla^2]+ \alpha_1(r)\nabla
+\alpha_2(r)\nabla^2\nonumber\\
&~&+\widetilde{V}(\vec{r})-E~)\Psi(\vec{r})=0
\end{eqnarray}
with
\begin{equation}
\nabla^2=\frac{1}{r}\frac{d^2}{dr^2}r-\frac{\overrightarrow{L}^2}{r^2},
\end{equation}
in which $\alpha(r)$,$\alpha_1(r)$ and $\alpha_2(r$) are
\begin{eqnarray}
\alpha(r)&=&(-2\mu)(-C_{\rho}\beta^2 g^2_v
\frac{\vec{\epsilon_b}\cdot \vec{\epsilon_a}^{\dag}}{2m m^*}
\mathcal{F}_{4t2 \rho}
-C_{\omega}\beta^2 g^2_v \frac{\vec{\epsilon_b}\cdot\vec{\epsilon_a}^{\dag}}{2m m^*}
\mathcal{F}_{4t2\omega}\nonumber\\
&~&+C C_{\rho}\lambda^2 g^2_v \frac{2(m^*-m)^2}{m m^*}
\vec{\epsilon_b}\cdot\vec{\epsilon_a}^{\dag}F_{4u2\rho}\nonumber\\
&~&+C C_{\omega}\lambda^2 g^2_v \frac{2(m^*-m)^2}{m m^*}
\vec{\epsilon_b}\cdot\vec{\epsilon_a}^{\dag}F_{4u2\omega})
\end{eqnarray}

\begin{eqnarray}
\alpha_1(r)&=&C C_{\pi}\frac{g^2_{\pi}}{f^2_{\pi}}\frac{(m^* -
m)^2}{m^{*2}}F'_{6u2\pi}\nonumber\\
&~&+C C_{\eta}\frac{g^2_{\eta}}{f^2_{\eta}}\frac{(m^* -
m)^2}{m^{*2}}F_{6u2\eta}\nonumber\\
&~&+C C_{\rho}\lambda^2 g^2_v \frac{2(2m^*+m)(m^* - m)^2}{m^{*3}}
F_{6u2\rho}\nonumber\\
&~&+C C_{\omega}\lambda^2 g^2_v \frac{2(2m^*+m)(m^* - m)^2}{m^{*3}}
F_{6u2\omega}
\end{eqnarray}

\begin{eqnarray}
\alpha_2(r)&=& C_{\pi}\frac{g^2_{\pi}}{f^2_{\pi}}\frac{(m^* -
m)^2}{m^{*2}}F'_{6u3\pi}\nonumber\\
&~&+C C_{\eta}\frac{g^2_{\eta}}{f^2_{\eta}}\frac{(m^* -
m)^2}{m^{*2}}F_{6u3\eta}\nonumber\\
&~&+C C_{\rho}\lambda^2 g^2_v \frac{2(2m^*+m)(m^* - m)^2}{m^{*3}}
F_{6u3\rho}\nonumber\\
&~&+C C_{\omega}\lambda^2 g^2_v \frac{2(2m^*+m)(m^* - m)^2}{m^{*3}}
F_{6u3\omega}
\end{eqnarray}

Then, with the wave function in Eq. (\ref{eq:tottrail}), the
Hamiltonian matrix can be expressed as
\begin{equation}
\begin{pmatrix}
H^{SS} & H^{SD}\\
H^{DS} & H^{DD}
\end{pmatrix}
\end{equation}
with
\begin{eqnarray}
H^{SS}&=&\langle \phi_S |\int^\infty _0 \sum^{n-1}_{i,j} a_i \chi_{i
0}(r) \{-\frac{\hbar^2}{2\mu}[1+\alpha(r)]\nabla^2 a_j \chi_{j 0}(r)\nonumber\\
&~&-\frac{\hbar^2}{2\mu}\nabla^2 [\alpha(r)a_j \chi_{j 0}(r)]
+\alpha_1(r)\nabla a_j \chi_{j 0}(r)\nonumber\\
&~&+\alpha_2(r)\nabla^2 a_j \chi_{j 0}(r) + V_{SS} (r) a_j \chi_{j
0}(r)\} r^2 dr | \phi_S \rangle,
\end{eqnarray}
\begin{eqnarray}
H^{SD}=\langle \phi_S |\int^\infty _0 \sum^{n-1}_{i,p} a_i \chi_{i
0}(r) V_{SD} (r) b_p \chi_{p 2}(r) r^2 dr | \phi_D \rangle,
\end{eqnarray}
\begin{eqnarray}
H^{DS}=\langle \phi_D |\int^\infty _0 \sum^{n-1}_{p,i} b_p \chi_{p
2}(r) V_{DS} (r)a_i \chi_{i 0}(r) r^2 dr | \phi_S \rangle,
\end{eqnarray}
and
\begin{eqnarray}
H^{DD}&=&\langle \phi_D |\int^\infty _0 \sum^{n-1}_{p,q} b_p \chi_{p
2}(r) \{-\frac{\hbar^2}{2\mu}[1+\alpha(r)]\nabla^2 b_q \chi_{q 2}(r)\nonumber\\
&~&-\frac{\hbar^2}{2\mu}\nabla^2 [\alpha(r)b_q \chi_{q 2}(r)]+\alpha_1(r)\nabla b_q \chi_{q 2}(r)\nonumber\\
&~&+\alpha_2(r)\nabla^2 b_q \chi_{q 2}(r) + V_{DD} (r) b_q \chi_{q
2}(r)\} r^2 dr  | \phi_D \rangle.
\end{eqnarray}
The total Hamiltonian contains three angular momentum related
operators $\hat{\vec{\epsilon_b}}\cdot
\hat{\vec{\epsilon_a}}^{\dag}$, $\hat{S}_{12}$,
$(\hat{\vec{\epsilon_b}}\times \hat{\vec{\epsilon_a}}^{\dag})\cdot
\hat{\vec{L}}$, which corresponds to the spin-spin interaction, the
spin orbit force and tensor force respectively. They act on the S
and D-wave coupled wave functions and split the total effective
potential $\widetilde{V}(\vec{r})$ into the subpotentials
$V_{SS}(r)$, $V_{SD}(r)$, $V_{DS}(r)$ and $V_{DD}(r)$. The matrix
form reads
\begin{eqnarray}
\langle \phi_S+\phi_D |\hat{\vec{\epsilon_b}}\cdot
\hat{\vec{\epsilon_a}}^{\dag}\widetilde{V}(\vec{r})|\phi_S+\phi_D\rangle=\left(
                                                \begin{array}{cc}
                                                  V_{SS}(r) & 0 \\
                                                  0 & V_{DD}(r) \\
                                                \end{array}
                                              \right)
\end{eqnarray}

\begin{eqnarray}
\langle \phi_S+\phi_D |\hat{S}_{12}\widetilde{V}(\vec{r})
|\phi_S+\phi_D\rangle=\left(\!
         \begin{array}{cc}
           \!\!0 & \!\!-\sqrt{2}V_{SD}(r) \\
           \!\!-\sqrt{2}V_{DS}(r) & \!\!1 \\
         \end{array}
      \! \right)
\end{eqnarray}

\begin{eqnarray}
\langle \phi_S+\phi_D |(\hat{\vec{\epsilon_b}}\times
\hat{\vec{\epsilon_a}}^{\dag})\cdot
\hat{\vec{L}}\widetilde{V}(\vec{r}) |\phi_S+\phi_D\rangle =\left(
                                                \begin{array}{cc}
                                                  0 & 0 \\
                                                  0 & 3i V_{DD}(r)\\
                                                \end{array}
                                              \right)
\end{eqnarray}
where the tensor force operator $\hat{S}_{12}$ mixes the S-wave and
D-wave contribution and is defined as
\begin{eqnarray}
\hat{S}_{12}=3(\vec{r}\cdot \hat{\vec{\epsilon_b}})(\vec{r}\cdot
\hat{\vec{\epsilon_a}}^{\dag})- \hat{\vec{\epsilon_b}}\cdot
\hat{\vec{\epsilon_a}}^{\dag}
\end{eqnarray}

\section{Numerical Results}\label{Numerical}

We diagonalize the Hamiltonian matrix to obtain the eigenvalue and
eigenvector. If there exists a negative eigenvalue, there exists a
bound state. The corresponding eigenvector is the wave function. We
use the variation principle to solve the equation. We change the
variable parameter to get the lowest eigenvalue. We also change the
number of the basis functions to reach a stable result.

\subsection{X(3872)}

The mass of the $\pi$ meson is smaller than the mass difference of
$D$ and $\bar{D}^*$, which causes the Fourier transformation of the
$\pi$-meson-exchange potential to be a complex function. The
different treatment of this complex potential would lead to quite
different results for the system. In our approach, we drop the the
imaginary part of the potential.

In order to distinguish each meson's effect, we plot each meson's
S-wave contribution to the potential in the first figure in
Fig.\ref{D-potential-meson}. The $\pi$ meson provides the most
attractive force while the $\sigma$ meson's attraction is relatively
small.

The main contribution to the binding energy comes form the S-wave
attractive force. We also plot the effective potential in the first
diagram in Fig. \ref{D-potential}. $V_s$ and $V_d$ are the effective
potential of the S-wave and D-wave interaction after adding the
momentum-related terms. $V'_s$ and $V'_d$ are the effective
potential of the S-wave and D-wave interaction without the
momentum-related terms. We can see a clear difference between $V_s$
and $V'_s$, which cause an obvious correction to the binding energy
when we consider the momentum-related terms.

We first used the computation programme to reproduce the deuteron
system successfully. Then we move on to investigate the possibility
of $X(3872)$ as the $D \bar{D}^{*}$ molecular state with quantum
number $I=0$, $J^{PC}=1^{++}$. For comparison, we first do not
consider the momentum-related terms. Then we add the
momentum-related terms and repeat the numerical analysis to
investigate its correction to the system.

Considering that the binding energy of $X(3872)$ is tiny, the
inclusion of the momentum-related terms may lead to significant
corrections to this very loosely bound system.

We collect the numerical results of the binding energy with the
variation of the cutoff parameter $\Lambda$ Table \ref{tab:X3872+}.
$E$ and $E'$ is the eigen-energy of Hamiltonian with and without the
momentum-related terms respectively. Besides the total energy, we
also list the separate contribution to the energy from the S-wave,
D-wave and spin-orbit force components respectively in the fourth,
fifth  and sixth column. The last column is the mass of X(3872) as a
molecular system.

There exists a bound state solution when the cutoff parameter
changes from $1.1 \sim 1.3 $ GeV. The binding energy with the recoil
correction is around $0.054 \sim 7.131$ MeV and the binding energy
without the recoil correction is around $0.276 \sim 9.686$ MeV. When
the binding energy is $7.131$ MeV with $\Lambda =1.3$ GeV, the
recoil correction is $2.555$ MeV and the contribution of the
spin-orbit force is $0.573$ MeV. When the binding energy is $2.361$
MeV, the recoil correction is $1.075$ MeV and the contribution of
the spin-orbit force is $0.213$ MeV. When the binding energy
decrease to $0.054$ with $\Lambda =1.1$ GeV, the recoil correction
reach $0.222$ MeV, which is even bigger than the binding energy
itself. Now the contribution of the spin-orbit force is $0.038$ MeV
and almost as big as the D-wave contribution. Clearly the recoil
correction decrease the binding energy and renders X(3872) to be an
extremely loosely bound molecular states partly.

\begin{table}[htbp]
\caption{The bound state solutions of the $D \bar{D}^{*}$ system
with $I^G=0^+$, $J^{PC}=1^{++}$ (in unit of MeV) with the cutoff
$\Lambda$. $E$ and $E'$ is the eigen-energy of the system with and
without the momentum-related terms respectively. We also list the
separate contribution to the energy from the S-wave, D-wave and
spin-orbit force components respectively in the fourth, fifth and
sixth column. The last column is the mass of X(3872) as a molecular
system.} \label{tab:X3872+}
\begin{center}
\begin{tabular}{c c c c c c | c}
\hline \hline \multirow{2}{*}{$\Lambda$(GeV)} & \multirow{2}{*}{~}  &\multicolumn{4}{c|}{Eigenvalue} & {Mass}  \\
\cline{2-6}
                                 &   &total   &S             &D        & LS   &(MeV) \\
\hline\hline
\multirow{2}{*}{1.10} & $E$    &  ~~-0.054~~  & ~~-4.364~~   & ~~0.052~~  & ~~0.038~~     & ~~3874.846~~\\
                      & $E'$   &  ~~-0.276~~  & ~~-4.458~~   & ~~0.017~~  & ~~-~~     & ~~3874.624~~\\
\hline\hline
\multirow{2}{*}{1.15} & $E$    &  ~~-0.884~~  & ~~-10.02~~   & ~~0.128~~   & ~~0.104~~   & ~~3874.016~~\\
                      & $E'$   &  ~~-1.449~~  & ~~-10.48~~   & ~~0.031~~   & ~~-~~   & ~~3873.451~~\\
\hline\hline
\multirow{2}{*}{1.20} & $E$    &  ~~-2.361~~  & ~~-17.23~~   & ~~0.245~~     & ~~0.213~~   & ~~3872.539~~\\
                      & $E'$   &  ~~-3.436~~  & ~~-18.14~~   & ~~0.046~~    & ~~-~~     & ~~3871.464~~\\
\hline\hline
\multirow{2}{*}{1.25} & $E$    &  ~~-4.469~~  & ~~-24.80~~    & ~~0.401~~    &  ~~0.367~~    & ~~3870.431~~\\
                       & $E'$  &  ~~-6.203~~  & ~~-26.29~~    & ~~0.059~~     & ~~-~~     & ~~3868.697~~ \\
\hline\hline
\multirow{2}{*}{1.30} & $E$    &  ~~-7.131~~  & ~~-32.45~~   & ~~0.609~~     & ~~0.573~~      & ~~3867.769~~\\
                       & $E'$  &  ~~-9.686~~  & ~~-34.63~~   & ~~0.076~~     &  ~~-~~   & ~~3865.214~~\\
\hline\hline
\end{tabular}
\end{center}
\end{table}

\subsection{The $D \bar{D}^{*}$ system with $I^G=0^-$,
$J^{PC}=1^{+-}$}

We also calculate the $D \bar{D}^{*}$ system with $I=0$,
$J^{PC}=1^{+-}$. The results with the variation of the cutoff from
$1.4 \sim 1.6 GeV$ are shown in the Table.\ref{tab:X3872-}. There
might also exist a bound state with odd C parity. Its binding energy
is slightly smaller than that of $X(3872)$ with the same cutoff.
When the binding energy is $2.386$ MeV with $\Lambda =1.4$ GeV, the
total recoil correction reaches $-0.447$ MeV while the contribution
of the spin-orbit force is $+0.9$ MeV, which is also almost as big
as the D-wave contribution. Clearly the recoil correction is
favorable to the formation of the molecular state in this channel.

The corresponding effective potential and the exchanged meson's
contribution are also shown in the second figure in
Fig.\ref{D-potential} and Fig.\ref{D-potential-meson}.

\begin{table}[htbp]
\caption{The bound state solution of the $D \bar{D}^{*}$ system with
$I^G=0^-$, $J^{PC}=1^{+-}$ (in unit of MeV) with $\Lambda$. $E$ and
$E'$ is the eigen-energy of the system with and without the
momentum-related terms respectively. We also list the separate
contribution to the energy from the S-wave, D-wave and spin-orbit
force components respectively in the fourth, fifth and sixth column.
The last column is the mass of the $D \bar{D}^{*}$ system with
$I^G=0^-$, $J^{PC}=1^{+-}$ as a molecular state.} \label{tab:X3872-}
\begin{center}
\begin{tabular}{c c c c c c | c }
\hline \hline \multirow{2}{*}{~$\Lambda$(GeV)~} & \multirow{2}{*}{~}  &\multicolumn{4}{c|}{Eigenvalue} & {Mass}   \\
\cline{2-6}
                                &   &total  &   S           &D              & LS          &(MeV) \\
\hline
\multirow{2}{*}{1.40} & $E$    &  ~~-2.386~~  & ~~-10.55~~ & ~~-1.587~~   & ~~0.900~~  & ~~3872.514~~ \\
                      & $E'$   &  ~~-1.939~~  & ~~-12.30~~ & ~~-2.371~~   & ~~-~~  & ~~3872.961~~\\
\hline\hline
\multirow{2}{*}{1.45} & $E$    &  ~~-7.098~~  & ~~-20.90~~ &~~-2.863~~    & ~~2.019~~   & ~~3867.802~~\\
                      & $E'$   &  ~~-6.298~~  & ~~-24.65~~ &~~-4.655~~    & ~~-~~ & ~~3868.602~~\\
\hline\hline
\multirow{2}{*}{1.50} & $E$    &  ~~-14.62~~  & ~~-33.88~~ &~~-4.236~~    & ~~3.635~~  & ~~3860.28~~\\
                      & $E'$   &  ~~-13.43~~  & ~~-40.41~~ & ~~-7.513~~   & ~~-~~ & ~~3861.47~~\\
\hline\hline
\multirow{2}{*}{1.55} & $E$    &  ~~-25.20~~  & ~~-49.63~~ &~~-5.657~~   & ~~5.822~~  & ~~3849.70~~\\
                      & $E'$  &  ~~-23.58~~   & ~~-59.87~~ &~~-10.97~~   & ~~-~~  & ~~3851.32~~\\
\hline\hline
\multirow{2}{*}{1.60} & $E$    &  ~~-39.10~~  & ~~-68.32~~ &~~-7.074~~   & ~~8.65~~  & ~~3835.80~~\\
                       & $E'$  &  ~~-36.95~~  & ~~-83.32~~ & ~~-15.06~~  & ~~-~~ & ~~3837.95~~\\
\hline\hline
\end{tabular}
\end{center}
\end{table}

\subsection{$Z_c(3900)$}

The newly observed $Z_c(3900)$ was explained as the isovector
partner of $X(3872)$ with $J^{PC}=1^{+-}$ by some theoretical groups
\cite{Q.Wang:2013, Z.G.Wang:2013, F.Aceti:2013}.

We carefully perform the investigation of the $D \bar{D}^{*}$ system
with $I^=1^+$, $J^{PC}=1^{+-}$. We consider the S-wave and D-wave
mixing, the spin orbit force at $O(1/M)$ and all the other possible
recoil corrections up to $O(1/M^2)$. The corresponding effective
potential and the exchanged meson's contribution are also shown in
the fourth diagram in Fig. \ref{D-potential} and Fig.
\ref{D-potential-meson}. Unfortunately, we are unable to obtain a
bound state solution with the pionic coupling $g=0.59 $ which was
extracted from the $D^*$ decay width. It seems there probably does
not exist a loosely bound isovector molecular state composed of the
$D \bar{D}^{*}$ mesons.

On the other hand, the $\pi$ meson exchange plays a dominant role.
Considering the uncertainty of $g$, we try to increase this coupling
constant to check the dependence of the results on $g$. We find when
the coupling constant $g$ increases by a factor of $1.6$ , a bound
state appears. The results are listed Table \ref{tab:Zc3900-}.

The binding energy of the $J^{PC}=1^{+-}$ molecule with/without the
recoil correction is around $0.037 \sim 15.82$ MeV and $0.322 \sim
18.51$ MeV respectively. When the binding energy is $0.037$ MeV, the
recoil correction is $0.285$ MeV and the contribution of the
spin-orbit force is $0.058$ MeV. Clearly the recoil corrections are
of the same order as the binding energy and unfavorable to the
formation of the molecular state.

\begin{table}[htbp]
\caption{The $I^G=1^+$, $J^{PC}=1^{+-}$ $D \bar{D}^{*}$ system with
with the enhanced coupling constant $g$ and $\Lambda = 2.0$ GeV. The
other notations are the same as in Table \ref{tab:X3872+}.}
\label{tab:Zc3900-}
\begin{center}
\begin{tabular}{c c c c c c | c}
\hline \hline \multirow{2}{*}{~$g\cdot n$~} & \multirow{2}{*}{~}  &\multicolumn{4}{c|}{Eigenvalue} & {Mass}   \\
\cline{2-6}
                                 &   &total  &   S          &D           & LS       &(MeV) \\
\hline\hline
\multirow{2}{*}{$g\cdot1.6$}       & $E$    &  ~~-0.037~~  & ~~-7.244~~ & ~~0.169~~ & ~~0.058~~ & ~~3874.863~~\\
                                   & $E'$   &  ~~-0.322~~  & ~~-7.421~~ &~~0.107~~ & ~~-~~ & ~~3874.578~~\\
\hline\hline
\multirow{2}{*}{$g\cdot1.7$}       & $E$    &  ~~-4.293~~  & ~~-42.32~~ &~~0.957~~ & ~~0.359~~ & ~~3870.607~~\\
                                   & $E'$   &  ~~-5.634~~  & ~~-43.21~~ &~~0.579~~ & ~~-~~ & ~~3869.266~~\\
\hline\hline
\multirow{2}{*}{$g\cdot1.8$}       & $E$    &  ~~-15.82~~  & ~~-93.30~~ &~~2.007~~ & ~~0.822~~ & ~~3859.08~~\\
                                   & $E'$   &  ~~-18.51~~  & ~~-95.04~~ & ~~1.146~~ & ~~-~~ & ~~3856.39~~\\
\hline\hline
\end{tabular}
\end{center}
\end{table}

\subsection{The $D \bar{D}^{*}$ system with $I^G=1^-$,
$J^{PC}=1^{++}$}

\begin{figure}[ht]
  \begin{center}
  \rotatebox{0}{\includegraphics*[width=0.38\textwidth]{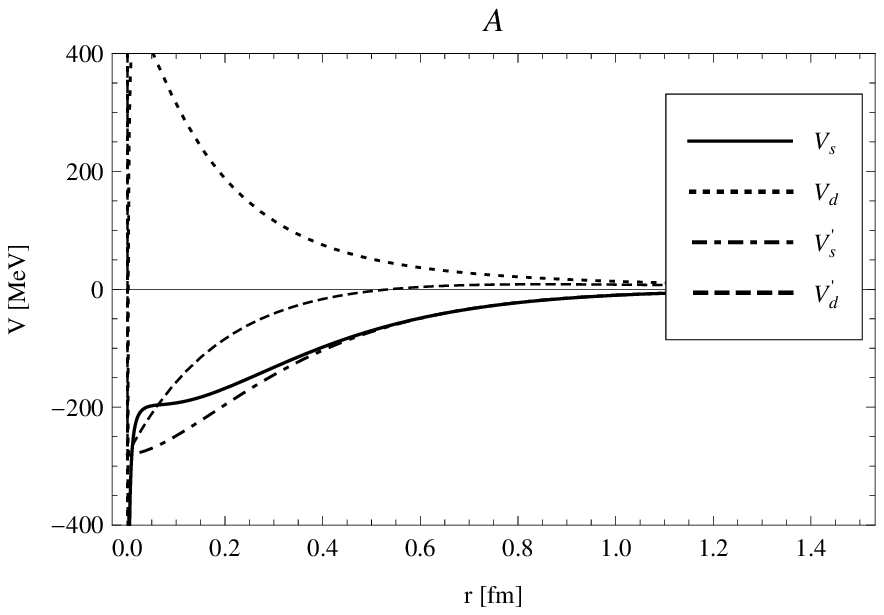}}
    \rotatebox{0}{\includegraphics*[width=0.38\textwidth]{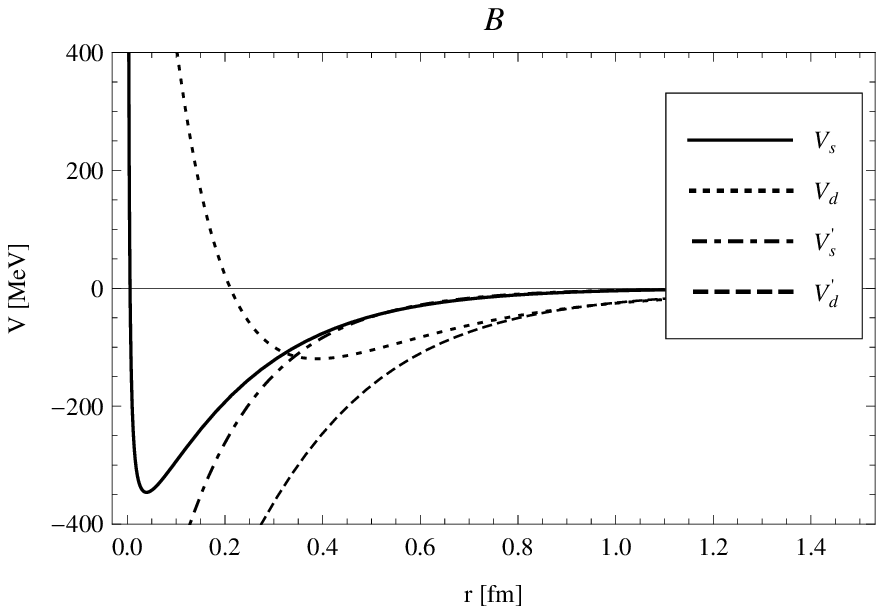}}
      \rotatebox{0}{\includegraphics*[width=0.38\textwidth]{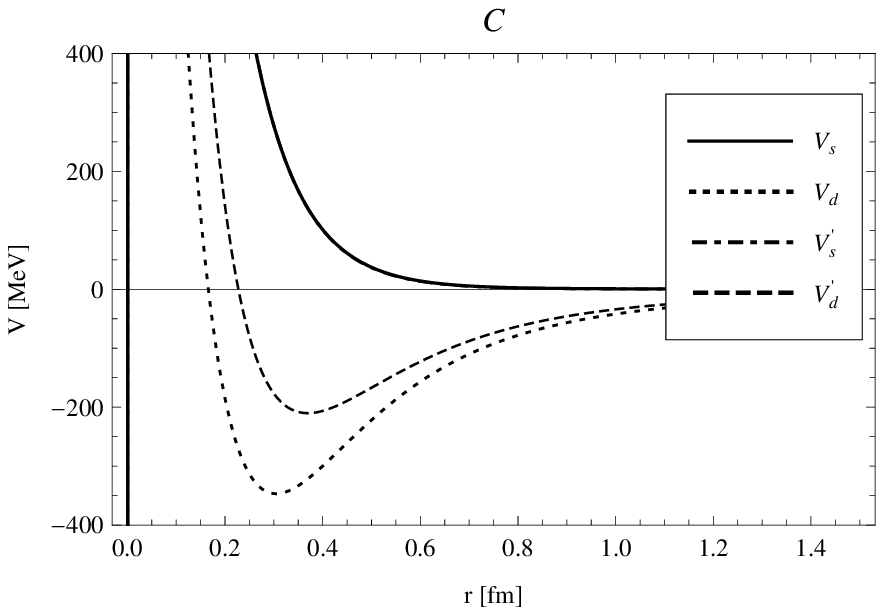}}
    \rotatebox{0}{\includegraphics*[width=0.38\textwidth]{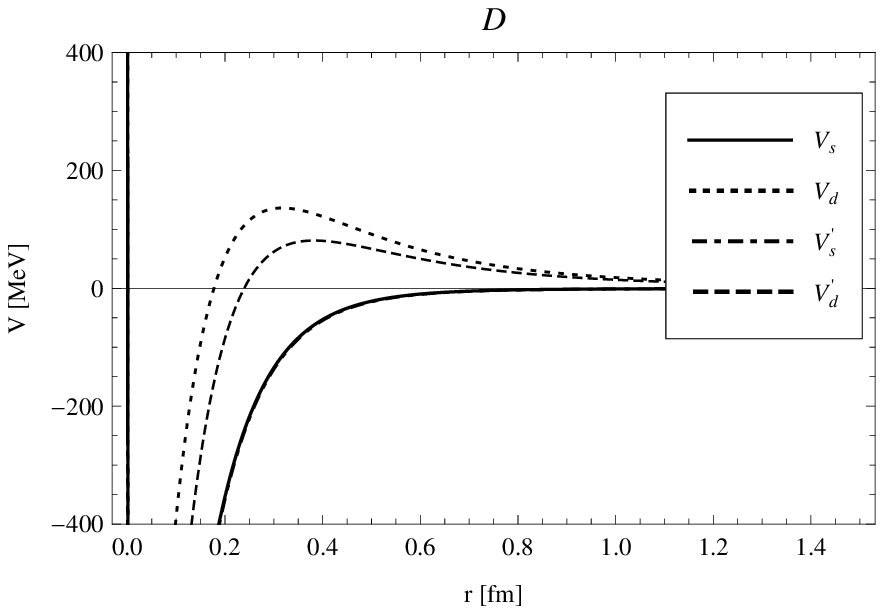}}
    \caption{The effective potential of the $D \bar{D}^{*}$ system.
    Labels A,B,C,D correspond to the four cases $I=0$, $J^{PC}=1^{++}$; $I=0$, $J^{PC}=1^{+-}$;
    $I=1$, $J^{PC}=1^{++}$; $I=1$, $J^{PC}=1^{+-}$ respectively
    from top to bottom. $V_s$ and $V_d$ are the effective potential of the S-wave and D-wave
    interaction with the momentum-related terms while $V'_s$ and $V'_d$ are
    the S-wave and D-wave effective potential without the momentum-related terms. }
    \label{D-potential}
  \end{center}
\end{figure}

\begin{figure}[ht]
  \begin{center}
  \rotatebox{0}{\includegraphics*[width=0.38\textwidth]{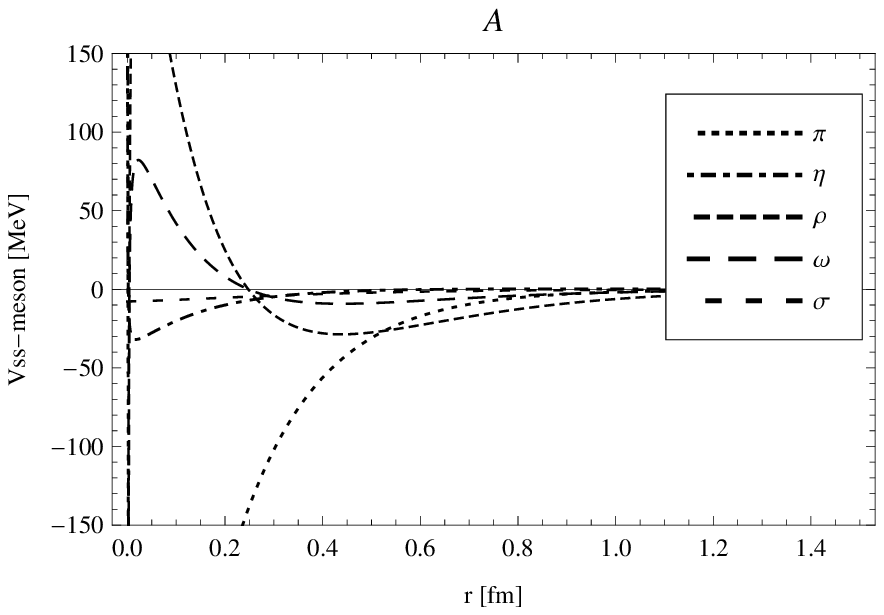}}
    \rotatebox{0}{\includegraphics*[width=0.38\textwidth]{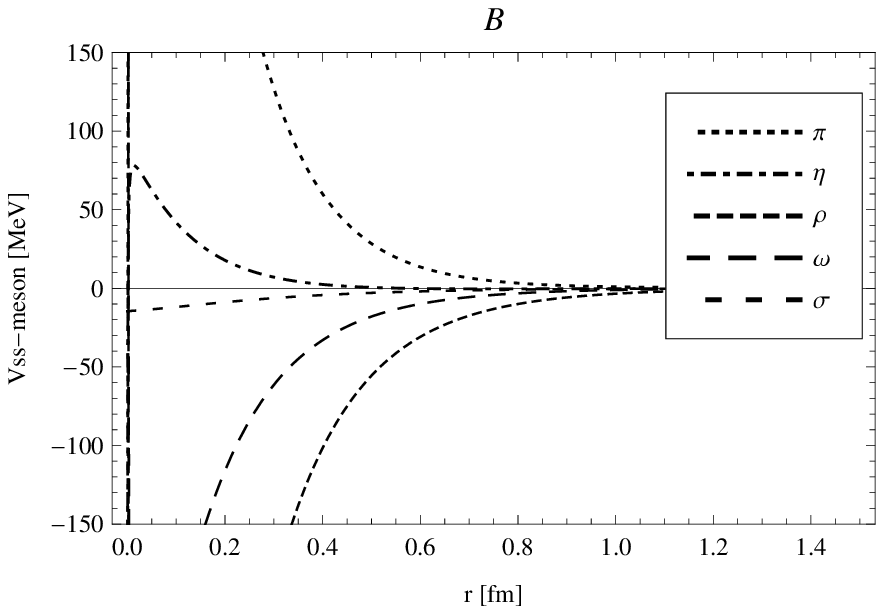}}
      \rotatebox{0}{\includegraphics*[width=0.38\textwidth]{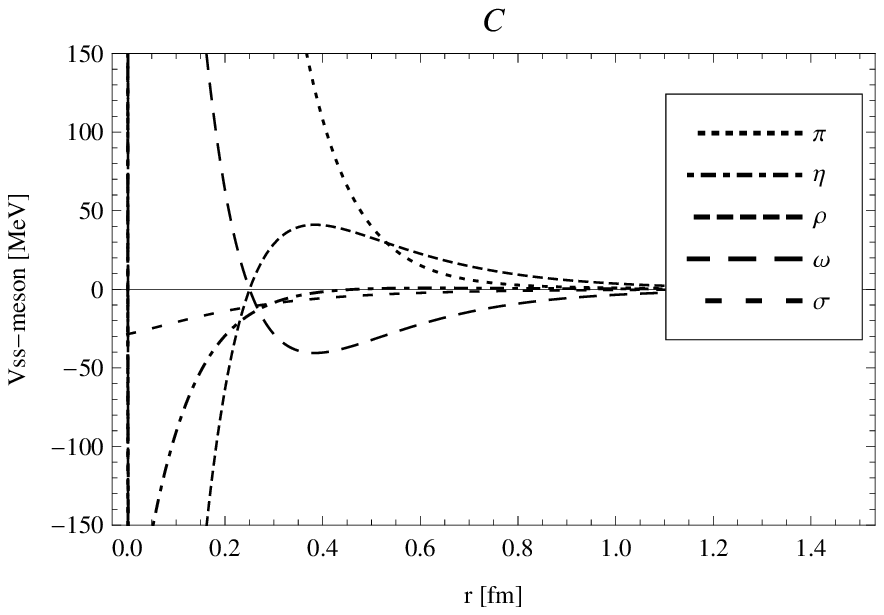}}
    \rotatebox{0}{\includegraphics*[width=0.38\textwidth]{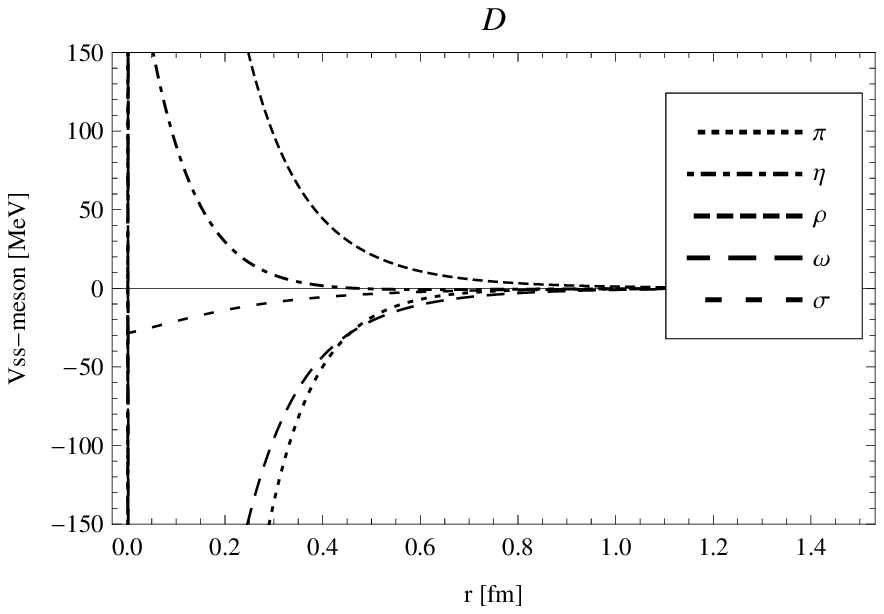}}
    \caption{The effective potential from the different meson exchange
    in the $D \bar{D}^{*}$ system. Labels A,B,C,D are the same as in
    Fig. \ref{D-potential}.}
    \label{D-potential-meson}
  \end{center}
\end{figure}

We also perform the investigation of the $D \bar{D}^{*}$ system with
$I^G=1^-$, $J^{PC}=1^{++}$. The corresponding effective potential
and the exchanged meson's contribution are also shown in the third
diagram in Fig. \ref{D-potential} and Fig. \ref{D-potential-meson}.
There does not exist a bound state solution with the pionic coupling
$g=0.59 $. If we increase $g$ by a factor $2.4$, there appears a
bound state. The numerical results are listed in Table
\ref{tab:Zc3900+}.

The binding energy of the possible $J^{PC}=1^{++}$ state with the
recoil correction is around $1.777 \sim 14.49$ MeV while it is
around $0.524 \sim 8.67$ MeV without the recoil correction. When the
binding energy is $1.777$ MeV, the total recoil correction is
$1.253$ MeV and the contribution of the spin-orbit force alone is
$-1.903$ MeV. The recoil corrections are comparable with the binding
energy and very favorable to the formation of the possible loosely
bound molecule.

\begin{table}[htbp]
\caption{The $I^G=1^-$, $J^{PC}=1^{++}$ $D \bar{D}^{*}$ system with
the enhanced coupling constant $g$ and $\Lambda =2$ GeV. The other
notations are the same as in Table \ref{tab:X3872+}. }
\label{tab:Zc3900+}
\begin{center}
\begin{tabular}{c c c c c c | c}
\hline \hline \multirow{2}{*}{~$g\cdot n$~} & \multirow{2}{*}{~}  &\multicolumn{4}{c|}{Eigenvalue} & {Mass}   \\
\cline{2-6}
                                   &   &total       &   S            &D              & LS          &(MeV) \\
\hline\hline
\multirow{2}{*}{$g\cdot2.4$}       & $E$    &  ~~-1.777~~  & ~~6.093~~ & ~~-7.074~~ & ~~-1.903~~ & ~~3873.123~~\\
                                   & $E'$   &  ~~-0.524~~  & ~~6.019~~ & ~~-5.202~~ & ~~-~~      & ~~3874.376~~\\
\hline\hline
\multirow{2}{*}{$g\cdot2.5$}       & $E$    &  ~~-6.518~~  & ~~12.23~~ & ~~-15.67~~ & ~~-4.313~~ & ~~3868.382~~\\
                                   & $E'$   &  ~~-3.311~~  & ~~12.11~~ & ~~-11.42~~ & ~~-~~      & ~~3871.589~~\\
\hline\hline
\multirow{2}{*}{$g\cdot2.6$}       & $E$    &  ~~-14.49~~  & ~~18.99~~ & ~~-26.35~~ & ~~-7.42~~  & ~~3860.41~~\\
                                   & $E'$   &  ~~-8.67~~   & ~~18.86~~ & ~~-19.02~~ & ~~-~~      & ~~3866.23~~\\
\hline\hline
\end{tabular}
\end{center}
\end{table}

\subsection{The $B \bar{B}^{*}$ system}

\begin{figure}[ht]
  \begin{center}
  \rotatebox{0}{\includegraphics*[width=0.38\textwidth]{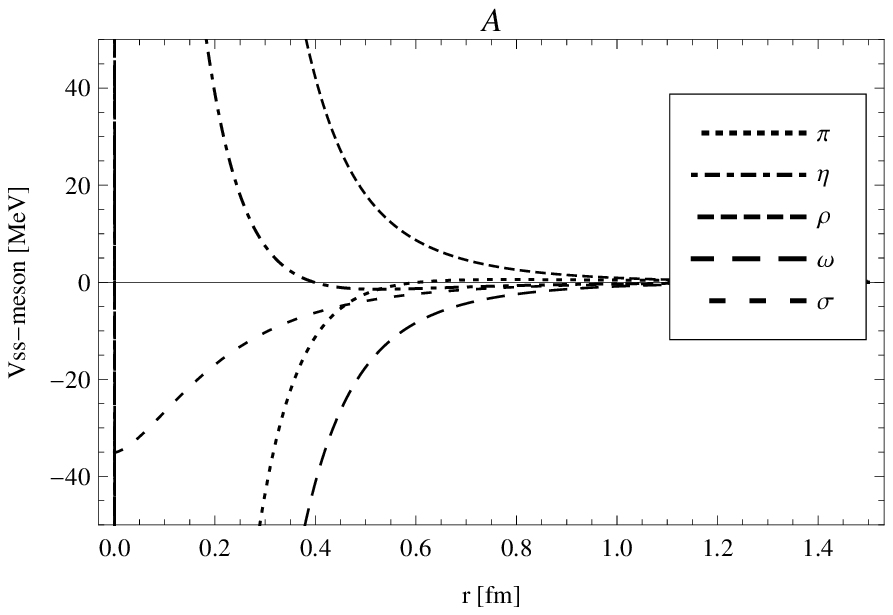}}
    \rotatebox{0}{\includegraphics*[width=0.38\textwidth]{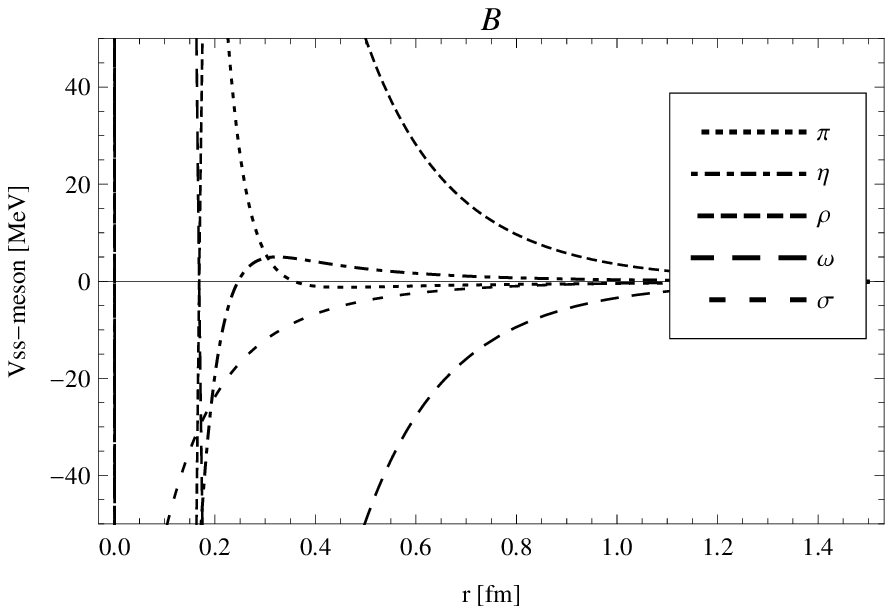}}
  \rotatebox{0}{\includegraphics*[width=0.38\textwidth]{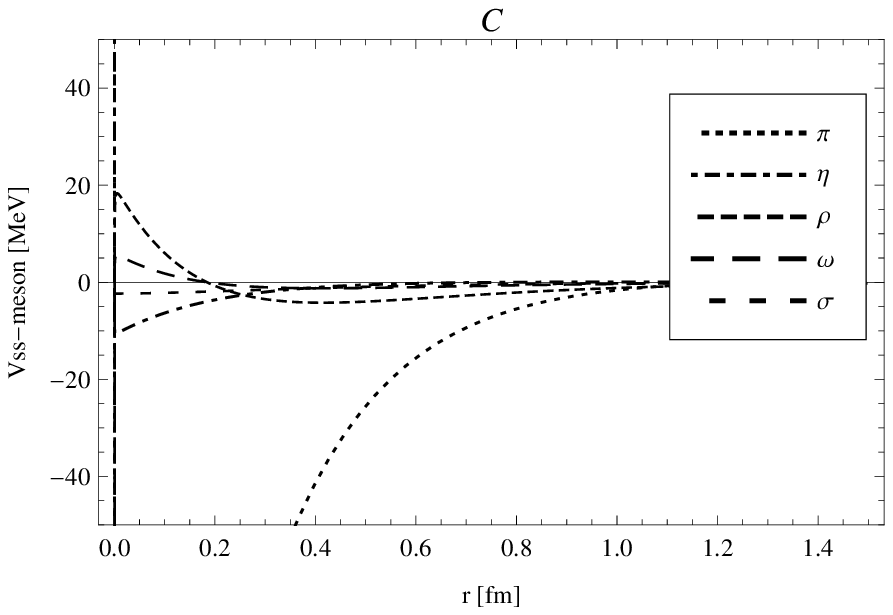}}
    \rotatebox{0}{\includegraphics*[width=0.38\textwidth]{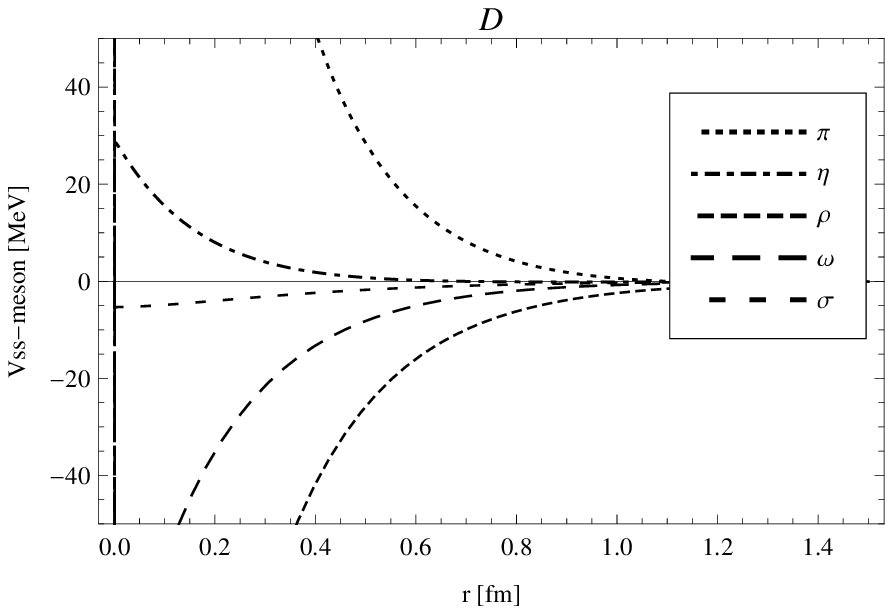}}
    \caption{The effective potential from the different meson exchange
   in the $B \bar{B}^{*}$ system. Labels A,B,C,D are the same as in
    Fig. \ref{D-potential}.}
    \label{B-potential-meson}
  \end{center}
\end{figure}

The effective potential and meson contributions are shown in Fig.
ref{B-potential}and Fig. \ref{B-potential-meson}. From Fig.
\ref{B-potential-meson}, one can see that the $\pi$ and $\rho$ and
$\omega$ mesons potentials are comparable.

\begin{figure}[ht]
  \begin{center}
  \rotatebox{0}{\includegraphics*[width=0.38\textwidth]{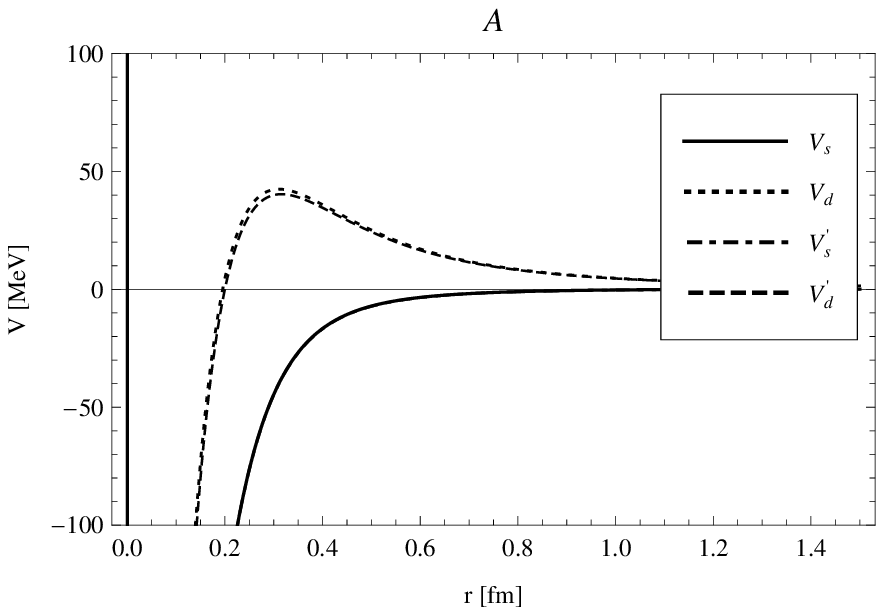}}
    \rotatebox{0}{\includegraphics*[width=0.38\textwidth]{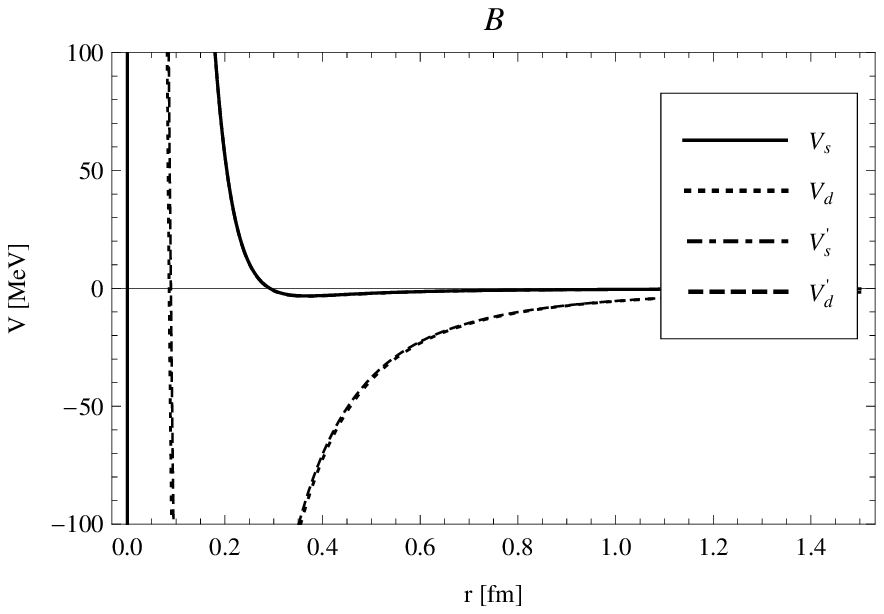}}
  \rotatebox{0}{\includegraphics*[width=0.38\textwidth]{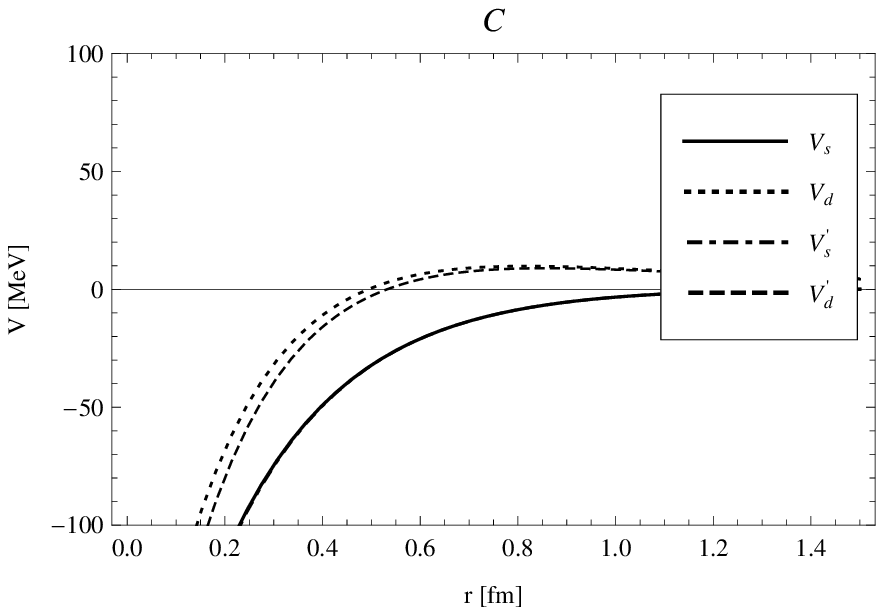}}
    \rotatebox{0}{\includegraphics*[width=0.38\textwidth]{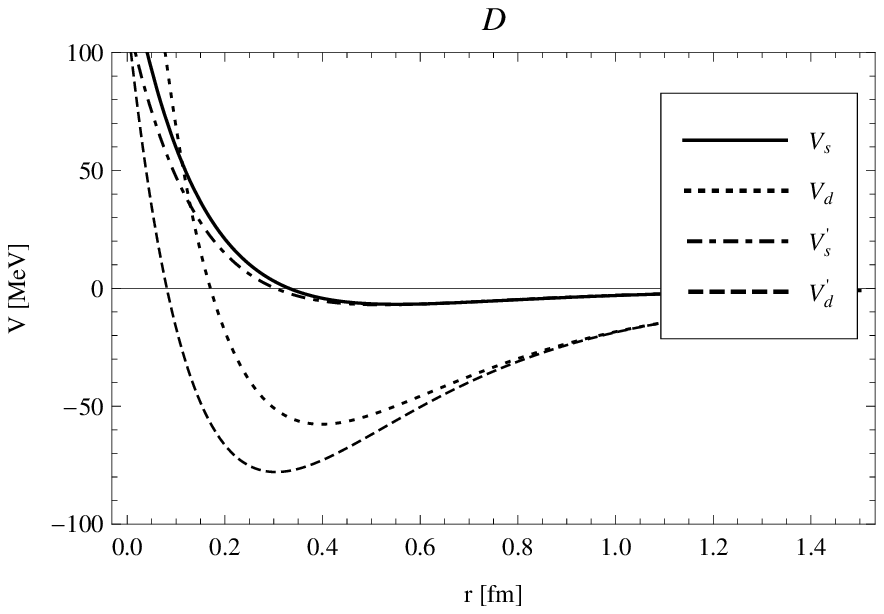}}
    \caption{ The effective potential of the $B \bar{B}^{*}$ system.
    Notations are the same as in Fig. \ref{D-potential}.
    }
    \label{B-potential}
  \end{center}
\end{figure}

Let's focus on the the momentum-related correction. From Fig.
\ref{B-potential}, we can see that the two curves of $V_s$ and
$V'_s$ almost overlap. The dominant momentum-related correction
comes from the D-wave interaction. In all cases, the
momentum-related correction is much smaller than that in the $D
\bar{D}^{*}$ system, which is expected because the $B$ meson is much
heavier than the $D$ meson.

For the $B \bar{B}^{*}$ system, there exist bound states with the
above three kinds of quantum number when varying the cutoff in an
appropriate range. We collect the numerical results in Tables
\ref{tab:Zb10610-}, \ref{tab:Zb10610-}, \ref{tab:BBstar+},
\ref{tab:BBstar-}.

The $I^G=1^+$, $J^{PC}=1^{+-}$ bound state corresponds to the
candidate of $Z_b(10610)$. The binding energy with the recoil
correction is around $0.251 \sim 18.5$ MeV and the binding energy
without recoil correction is about $0.348 \sim 19.58$ MeV with the
cutoff from $2.1 \sim 2.9$ GeV. When the binding energy is $0.251$
MeV, the recoil correction is $0.097$ MeV.

For the $I^G=1^+$, $J^{PC}=1^{+-}$ bound state, its binding energy
with the recoil correction is around $0.02 \sim 0.446$ MeV and about
$0.065 \sim 0.56$ MeV without the recoil correction with the cutoff
varies from $4.9 \sim 5.1$ GeV. However, this cutoff may be too
larger for a loosely bound system. Its binding energy is much
smaller than that of $Z_b(10610)$. When the binding energy is $0.02$
MeV, the recoil correction is $0.045$ MeV and the contribution of
spin-orbit force is $0.04$ MeV.

There exist two $I=0$ bound states which might be the isocalar
partners of $Z_b(10610)$. For the $I^G=0^+$, $J^{PC}=1^{++}$
molecule, the binding energy with the recoil correction is about
$0.28 \sim 36.87$ MeV when the cutoff varies from $0.7 \sim 1.1$
GeV. When the binding energy is $0.28$ MeV, the recoil correction is
$0.047$ MeV. For the $I^G=0^-$, $J^{PC}=1^{+-}$ molecule, the
binding energy with the recoil correction varies from $0.29 \sim
21.09$ MeV with the cutoff around $1.0 \sim 1.2$ GeV.

\begin{table}[htbp]
\caption{The $B \bar{B}^{*}$ system with $I^G=1^+$, $J^{PC}=1^{+-}$
(in unit of MeV). The other notations are the same as in Table
\ref{tab:X3872+}.} \label{tab:Zb10610-}
\begin{center}
\begin{tabular}{c c c c c c | c}
\hline \hline \multirow{2}{*}{~$\Lambda (GeV)$~} & \multirow{2}{*}{~}  &\multicolumn{4}{c|}{Eigenvalue} & {Mass}   \\
\cline{2-6}
                                 &   &total  &   S  &D   & LS &($MeV$) \\
\hline\hline
\multirow{2}{*}{2.1}  & $E$    &  ~~-0.251~~  & ~~-6.320~~ & ~~0.079~~ & ~~0.0008~~& ~~10603.749~~\\
                      & $E'$   &  ~~-0.348~~  & ~~-6.337~~ &~~0.075~~ & ~~-~~& ~~10603.652~~\\
\hline\hline
\multirow{2}{*}{2.3}  & $E$    &  ~~-1.766~~  & ~~-18.76~~ &~~0.227~~ & ~~0.011~~& ~~10602.234~~\\
                      & $E'$   &  ~~-2.026~~  & ~~-19.11~~ &~~0.214~~ & ~~-~~& ~~10601.974~~\\
\hline\hline
\multirow{2}{*}{2.5}  & $E$    &  ~~-4.988~~  & ~~-36.17~~ &~~0.430~~ & ~~0.022~~& ~~10599.012~~\\
                      & $E'$   & ~~-5.461~~   & ~~-36.21~~ &~~0.404~~ & ~~-~~& ~~10598.539~~\\
\hline\hline
\multirow{2}{*}{2.7}  & $E$    &  ~~-10.39~~  & ~~-59.52~~ &~~0.706~~ & ~~0.038~~& ~~10593.61~~\\
                      & $E'$   &  ~~-11.14~~  & ~~-59.56~~ &~~0.663~~ & ~~-~~& ~~10592.86~~\\
\hline\hline
\multirow{2}{*}{2.9}  & $E$    &  ~~-18.50~~  & ~~-89.71~~ &~~1.075~~ & ~~0.058~~& ~~10585.50~~\\
                      & $E'$   &  ~~-19.58~~  & ~~-89.74~~ &~~1.009~~ & ~~-~~& ~~10584.42~~\\
\hline\hline
\end{tabular}
\end{center}
\end{table}

\begin{table}[htbp]
\caption{The $B \bar{B}^{*}$ system with $I^G=1^-$, $J^{PC}=1^{++}$
(in unit of MeV). The other notations are the same as in Table
\ref{tab:X3872+}.} \label{tab:Zb10610+}
\begin{center}
\begin{tabular}{c c c c c c | c}
\hline \hline \multirow{2}{*}{~$\Lambda (GeV)$~} & \multirow{2}{*}{~}  &\multicolumn{4}{c|}{Eigenvalue} & {Mass}   \\
\cline{2-6}
                     &   &total&   S  &D   & LS      &($MeV$) \\
\hline\hline
\multirow{2}{*}{4.9}  & $E$    &  ~~-0.02~~   & ~~0.772~~ & ~~-0.932~~ & ~~-0.040~~ & ~~10603.98~~\\
                      & $E'$   &  ~~-0.065~~  & ~~0.764~~ &~~-0.895~~ & ~~-~~       & ~~10603.935~~\\
\hline\hline
\multirow{2}{*}{4.95} & $E$    &  ~~-0.089~~  & ~~1.049~~ &~~-1.252~~ & ~~-0.054~~ & ~~10603.911~~\\
                      & $E'$   &  ~~-0.148~~  & ~~1.039~~ &~~-1.202~~ & ~~-~~      & ~~10603.852~~\\
\hline\hline
\multirow{2}{*}{5.0}  & $E$    &  ~~-0.18~~  & ~~1.397~~ &~~-1.665~~ & ~~-0.072~~ & ~~10603.820~~\\
                      & $E'$   &  ~~-0.256~~  & ~~1.384~~ & ~~-1.598~~ & ~~-~~    & ~~10603.744~~\\
\hline\hline
\multirow{2}{*}{5.05} & $E$    &  ~~-0.298~~  & ~~1.809~~ &~~-2.155~~ & ~~-0.094~~& ~~10603.702~~\\
                       & $E'$  &  ~~-0.392~~   & ~~1.792~~ & ~~-2.068~~ & ~~-~~& ~~10603.608~~\\
\hline\hline
\multirow{2}{*}{5.1}  & $E$    &  ~~-0.446~~  & ~~2.273~~ &~~-2.710~~ & ~~-0.118~~& ~~10603.554~~\\
                       & $E'$  &  ~~-0.56~~  & ~~2.254~~ & ~~-2.60~~ & ~~-~~& ~~10603.440~~\\
\hline\hline
\end{tabular}
\end{center}
\end{table}

\begin{table}[htbp]
\caption{The $B \bar{B}^{*}$ system with $I^G=0^+$, $J^{PC}=1^{++}$
(in unit of MeV). The other notations are the same as in Table
\ref{tab:X3872+}.} \label{tab:BBstar+}
\begin{center}
\begin{tabular}{c c c c c c | c}
\hline \hline \multirow{2}{*}{~$\Lambda (GeV)$~} & \multirow{2}{*}{~}  &\multicolumn{4}{c|}{Eigenvalue} & {Mass}   \\
\cline{2-6}
                                 &   &total&   S  &D  & LS   &($MeV$) \\
\hline
\multirow{2}{*}{0.7}  & $E$     &  ~~-0.280~~  & ~~-3.174~~ &~~0.039~~ & ~~0.005~~ & ~~10603.720~~\\
                      & $E'$    &  ~~-0.327~~  & ~~-3.178~~ &~~0.034~~ & ~~-~~ & ~~10603.673~~\\
\hline\hline
\multirow{2}{*}{0.8}  & $E$    &  ~~-0.930~~   & ~~-6.631~~ &~~0.108~~ & ~~0.008~~ & ~~10603.070~~\\
                      & $E'$   &  ~~-1.027~~   & ~~-6.615~~ &~~0.100~~ & ~~-~~ & ~~10602.973~~\\
\hline\hline
\multirow{2}{*}{0.9}  & $E$    &  ~~-6.631~~  & ~~-22.31~~ &~~0.188~~ & ~~0.050~~ & ~~10597.369~~\\
                      & $E'$   &  ~~-7.705~~  & ~~-22.45~~ &~~0.140~~ & ~~-~~ & ~~10596.295~~\\
\hline\hline
\multirow{2}{*}{1.0}  & $E$    &  ~~-19.08~~  & ~~-44.46~~ &~~0.034~~ & ~~0.206~~ & ~~10584.920~~\\
                       & $E'$  &  ~~-20.42~~   & ~~-45.08~~ &~~0.663~~ & ~~-~~ & ~~10583.58~~\\
\hline\hline
\multirow{2}{*}{1.1}   & $E$    &  ~~-36.87~~  & ~~-67.91~~ &~~0.403~~ & ~~0.590~~ & ~~10567.13~~\\
                       & $E'$   & ~~-39.87~~  & ~~-69.45~~  &~~-0.158~~ & ~~-~~ & ~~10643.87~~\\
\hline\hline
\end{tabular}
\end{center}
\end{table}

\begin{table}[htbp]
\caption{The $B \bar{B}^{*}$ system with $I^G=0^-$, $J^{PC}=1^{+-}$
(in unit of MeV). The other notations are the same as in Table
\ref{tab:X3872+}.} \label{tab:BBstar-}
\begin{center}
\begin{tabular}{c c c c c c| c }
\hline \hline \multirow{2}{*}{~$\Lambda (GeV)$~} & \multirow{2}{*}{~}  &\multicolumn{4}{c|}{Eigenvalue}  & {Mass}  \\
\cline{2-6}
                                &   &total&   S  &D  &  LS &($MeV$) \\
\hline\hline
\multirow{2}{*}{1.0}  & $E$     &  ~~-0.290~~  & ~~-0.042~~ & ~~-0.568~~ & ~~0.023~~& ~~10603.710~~\\
                      & $E'$    &  ~~-0.290~~  & ~~-0.049~~ &~~-0.584~~  & ~~-~~& ~~10603.710~~\\
\hline\hline
\multirow{2}{*}{1.05}  & $E$    &  ~~-1.838~~   & ~~-0.502~~ &~~-1.832~~ & ~~0.132~~& ~~10602.162~~\\
                      & $E'$    &  ~~-1.841~~   & ~~-0.548~~ &~~-1.936~~ & ~~-~~& ~~10602.159~~\\
\hline\hline
\multirow{2}{*}{1.1}  & $E$    &  ~~-5.388~~  & ~~-1.992~~ &~~-3.794~~ & ~~0.409~~& ~~10598.612~~\\
                      & $E'$   &  ~~-5.439~~  & ~~-2.145~~ &~~-4.135~~ & ~~-~~& ~~10598.561~~\\
\hline\hline
\multirow{2}{*}{1.15}  & $E$    &  ~~-11.60~~  & ~~-5.079~~ &~~-6.447~~ & ~~0.939~~& ~~10592.40~~\\
                       & $E'$   &  ~~-11.81~~  & ~~-5.451~~ &~~-7.256~~ & ~~-~~& ~~10592.19~~\\
\hline\hline
\multirow{2}{*}{1.2}   & $E$    &  ~~-21.02~~  & ~~-10.24~~ &~~9.779~~  & ~~1.798~~& ~~10582.98~~\\
                       & $E'$   &  ~~-21.59~~  & ~~-10.99~~ &~~-11.37~~ & ~~-~~& ~~10582.41~~\\
\hline\hline
\end{tabular}
\end{center}
\end{table}

\section{Summary and Discussion}\label{summary}

In the framework of the one boson exchange model, we have calculated
the effective potentials between two heavy mesons from the t- and
u-channel $\pi$, $\eta$, $\rho$, $\omega$ and $\sigma$ meson
exchange. We keep the recoil corrections to the $B \bar{B}^{*}$ and
$D \bar{D}^{*}$ system up to $O(\frac{1}{M^2})$. We also keep terms
related to $\vec{k}=\frac{1}{2}(\vec{p'}+\vec{p})$, which is the sum
of the initial and final momentum of the system. Especially, the
spin orbit force appears at $O(\frac{1}{M})$, which turns out to be
important for the very loosely bound molecular states.

We have carefully investigated the $B \bar{B}^{*}$ and $D
\bar{D}^{*}$ systems with four kinds of quantum number: $I=0$,
$J^{PC}=1^{++}$; $I=0$, $J^{PC}=1^{+-}$; $I=1$, $J^{PC}=1^{++}$;
$I=1$, $J^{PC}=1^{+-}$.

After solving the Schr$\ddot{o}$dinger equation by the variation
method, we notice that there exist two isoscalar $D \bar{D}^{*}$
molecular states with $J^{PC}=1^{++}$ and $J^{PC}=1^{+-}$ with or
without the momentum-related corrections. The first C-parity even
$1^{++}$ state corresponds to $X(3872)$. In contrast, there exist
loosely bound states in three channels for the $B \bar{B}^{*}$
system with the cutoff in a reasonable range.

Our numerical results show that the momentum-related corrections are
unfavorable to the formation of the loosely bound molecular states
in the $I=0$, $J^{PC}=1^{++}$ and $I=1$, $J^{PC}=1^{+-}$ channels in
the $D \bar{D}^{*}$ system. Especially the recoil corrections are
quite large. For example, the recoil correction may be larger than
the binding energy of X(3872), which may partly force X(3872) to
become a very shallow bound state. As expected, the recoil
correction in the $D \bar{D}^{*}$ system is much larger than that in
the $B \bar{B}^{*}$ system.

However, we are unable to find a bound state for the $D \bar{D}^{*}$
system with $I=1$, $J^{PC}=1^{++}$ and $I=1$, $J^{PC}=1^{+-}$ with
the pionic coupling $g=0.59 $ which was extracted from the $D^*$
decay width and plays a dominant role in the effective potential,
although we have systematically included the S-D wave mixing effect,
the spin orbit force and all the other recoil corrections up to
$O(\frac{1}{M^2})$. A loosely bound state appears if we increase $g$
manually by a factor of $1.6\sim 1.8$ after the inclusion of the
recoil corrections.

It seems that it's not so easy to accommodate the newly observed
charged resonance $Z_c(3900)$ as the candidate of the isovector
molecular state of $D \bar{D}^{*}$. The present investigation shows
that the recoil corrections may diminish the binding energy by one
to several MeV and are unfavorable to the formation of loosely bound
molecular states in this channel. Experimentally the mass
$Z_c(3900)$ seems above the $D \bar{D}^{*}$ threshold. Our analysis
shows that there does exist attraction in this channel. One may
wonder whether $Z_c(3900)$ is a candidate of the molecular-type
resonance instead of a $D \bar{D}^{*}$ bound state.

On the other hand, we should also inspect the framework of the one
boson exchange model. One obvious uncertainty arises from the cutoff
parameter, which is commonly used to suppress the high momentum
contribution. Moreover, in the derivation of the effective
potential, we make Fourier transformation to the effective potential
in the momentum space to derive the potential in the coordinate
space. In case of the $D \bar{D}^{*}$ system, the mass splitting
between $D $ and $ \bar{D}^{*}$ is larger than the pion mass. Hence
the integral contains an imaginary part. The commonly used approach
is to take the principal value of this integral and omit the
imaginary part in order to ensure the effective potential and
Hamiltonian to be real. The resulting potential is oscillating. The
reliability of such a formalism deserves further investigation.

In short, the XYZ states provide a unique platform to study the
complicated low energy strong dynamics. The charmonium (or Upsilon)
spectrum above the open charm (or bottom) threshold and those
charmonium-like XYZ states as non-conventional candidates are
particularly interesting. In order to interpret their underlying
structures, we need also investigate their decay pattern and
production mechanisms.

\section{acknowledgement}
We thank Li-Ping Sun, Zhi-Feng Sun and Li Ning for useful
discussions. This project is supported by the National Natural
Science Foundation of China under Grant No. 11261130311.

\section{Appendix}

We collect the expressions of the functions used in the previous
sections in the appendix.
\begin{equation}
Y(\tilde{m}_{\alpha}r)=\frac{\exp(\tilde{m}_{\alpha}r)}{\tilde{m}_{\alpha}r}
\end{equation}
\begin{equation}
Z(\tilde{m}_{\alpha}r)=(1+\frac{3}{\tilde{m}_{\alpha}r}+\frac{3}{(\tilde{m}_{\alpha}r)^2})Y(\tilde{m}_{\alpha}r)
\end{equation}
\begin{equation}
Z_1(\tilde{m}_{\alpha}r)=(\frac{1}{\tilde{m}_{\alpha}r}+\frac{1}{(\tilde{m}_{\alpha}r)^2})Y(\tilde{m}_{\alpha}r)
\end{equation}
\begin{equation}
Z'(\tilde{m}_{\alpha}r)=\frac{\sin(\tilde{m}_{\alpha}r)}{\tilde{m}_{\alpha}r}-\frac{3}{\tilde{m}_{\alpha}r}\frac{\sin(\tilde{m}_{\alpha}r)}{\tilde{m}_{\alpha}r}+\frac{1}{(\tilde{m}_{\alpha}r)^2}\frac{\cos(\tilde{m}_{\alpha}r)}{\tilde{m}_{\alpha}r}.
\end{equation}
\begin{equation}
Z'_1(\tilde{m}_{\alpha}r)=\frac{1}{\tilde{m}_{\alpha}r}\frac{\sin(\tilde{m}_{\alpha}r)}{\tilde{m}_{\alpha}r}+\frac{1}{(\tilde{m}_{\alpha}r)^2}\frac{\cos(\tilde{m}_{\alpha}r)}{\tilde{m}_{\alpha}r}
\end{equation}
where for the $D \bar{D}^*$ system
\begin{equation}
\tilde{m}^2_{\pi}=(m_D^*-m_D)^2-m^2_{\pi},
\end{equation}
\begin{equation}
\tilde{m}^2_{\sigma,\rho,\omega,\eta}=m^2_{\sigma,\rho,\omega,\eta}-(m_D^*-m_D)^2.
\end{equation}
whille for the $B \bar{B}^*$ system
\begin{equation}
\tilde{m}^2_{\pi,\sigma,\rho,\omega,\eta}=m^2_{\pi,\sigma,\rho,\omega,\eta}-(m_B^*-m_B)^2.
\end{equation}
\begin{eqnarray}
\mathcal{F}_{1t\alpha}&=&\mathcal{F}\{(\frac{\Lambda^2-m_{\alpha}^2}{{\Lambda}^2+\vec{q}^2})
\frac{1}{\vec{q}^2+m_{\alpha}^2}\}\nonumber\\
&=&m_{\alpha}Y(m_{\alpha}r)-\Lambda Y(\Lambda r)-
(\Lambda^2-m_{\alpha}^2)\frac{e^{-\Lambda r}}{2\Lambda}
\end{eqnarray}
\begin{eqnarray}
\mathcal{F}_{1u\alpha}&=&\mathcal{F}\{(\frac{\Lambda^2-m_{\alpha}^2}{\tilde{\Lambda}^2+\vec{q}^2})
\frac{1}{\vec{q}^2+\tilde{m}_{\alpha}^2}\}\nonumber\\
&=&\tilde{m}_{\alpha}Y(\tilde{m}_{\alpha}r)-\tilde{\Lambda}
Y(\tilde{\Lambda}
r)-(\Lambda^2-m_{\alpha}^2)\frac{e^{-\tilde{\Lambda}
r}}{2\tilde{\Lambda}}
\end{eqnarray}

\begin{eqnarray}
\mathcal{F}_{2t\alpha}&=&\mathcal{F}\{(\frac{\Lambda^2-m_{\alpha}^2}{{\Lambda}^2+\vec{q}^2})
\frac{\vec{q}^2}{\vec{q}^2+m_{\alpha}^2}\}\nonumber\\
&=&m_{\alpha}^2[\Lambda Y(\Lambda r)-m_{\alpha}Y(m_{\alpha}r)]\nonumber\\
&+&(\Lambda^2-m_{\alpha}^2)\Lambda\frac{e^{-\Lambda r}}{2}
\end{eqnarray}

\begin{eqnarray}
\mathcal{F}_{2u\alpha}&=&\mathcal{F}\{(\frac{\Lambda^2-m_{\alpha}^2}{{\tilde{\Lambda}}^2+\vec{q}^2})
\frac{\vec{q}^2}{\vec{q}^2+\tilde{m}_{\alpha}^2}\}\nonumber\\
&=&\tilde{m}_{\alpha}^2[\tilde{\Lambda} Y(\tilde{\Lambda} r)-\tilde{m}_{\alpha}Y(\tilde{m}_{\alpha}r)]\nonumber\\
&+&(\Lambda^2-m_{\alpha}^2)\tilde{\Lambda}\frac{e^{-\tilde{\Lambda}
r}}{2}
\end{eqnarray}

\begin{eqnarray}
\mathcal{F}_{3t\alpha} &=&
\mathcal{F}\{(\frac{\Lambda^2-m_{\alpha}^2}{{\Lambda}^2+\vec{q}^2})
\frac{(\vec{\sigma_1}\cdot\vec{q})(\vec{\sigma_2}\cdot\vec{q})}{\vec{p}^2+m_{\alpha}^2}\}\nonumber\\
&=&\frac{1}{3}\vec{\sigma_1}\cdot\vec{\sigma_2}[~m_{\alpha}^2\Lambda Y(\Lambda r)-m_{\alpha}^3 Y(m_{\alpha}r)\nonumber\\
&+&(\Lambda^2-m_{\alpha}^2)\Lambda\frac{e^{-\Lambda r}}{2}~]\nonumber\\
&+&\frac{1}{3}S_{12}[-m_{\alpha}^3 Z(m_{\alpha}r)+ \Lambda^3 Z(\Lambda r) \nonumber\\
&+&(\Lambda^2-m_{\alpha}^2)(1+\Lambda r)
\frac{\Lambda}{2}Y(\Lambda r)]\nonumber\\
&=&(\vec{\sigma_1}\cdot\vec{\sigma_2})\mathcal{F}_{3t1} +
S_{12}\mathcal{F}_{3t2}
\end{eqnarray}

\begin{eqnarray}
\mathcal{F}_{3u\alpha} &=&
\mathcal{F}\{(\frac{\Lambda^2-m_{\alpha}^2}{{\tilde{\Lambda}}^2+\vec{q}^2})
\frac{(\vec{\sigma_1}\cdot\vec{q})(\vec{\sigma_2}\cdot\vec{q})}{\vec{q}^2+\tilde{m}_{\alpha}^2}\}\nonumber\\
&=&\frac{1}{3}\vec{\sigma_1}\cdot\vec{\sigma_2}[\tilde{m}_{\alpha}^2\tilde{\Lambda} Y(\tilde{\Lambda} r)-\tilde{m}_{\alpha}^3Y(\tilde{m}_{\alpha}r)\nonumber\\
&+&(\Lambda^2-m_{\alpha}^2)\tilde{\Lambda}\frac{e^{-\tilde{\Lambda} r}}{2}]\nonumber\\
&+&\frac{1}{3}S_{12}[-\tilde{m_{\alpha}}^3 Z(\tilde{m_{\alpha}}r)+ \tilde{\Lambda}^3 Z(\tilde{\Lambda} r) \nonumber\\
&+&(\Lambda^2-m_{\alpha}^2)(1+\tilde{\Lambda} r)\frac{\tilde{\Lambda}}{2}Y(\tilde{\Lambda} r)~]\nonumber\\
&=&(\vec{\sigma_1}\cdot\vec{\sigma_2})\mathcal{F}_{3u1\alpha} +
S_{12}\mathcal{F}_{3u2\alpha}
\end{eqnarray}

\begin{eqnarray}
\mathcal{F'}_{3u\alpha}&=&
\mathcal{F}\{(\frac{\Lambda^2-m_{\alpha}^2}{{\tilde{\Lambda}}^2+\vec{q}^2})
\frac{(\vec{\sigma_1}\cdot\vec{q})(\vec{\sigma_2}\cdot\vec{q})}{\vec{p}^2-\tilde{m_{\alpha}}^2}\}\nonumber\\
&=&\frac{1}{3}\vec{\sigma_1}\cdot\vec{\sigma_2}[~-\tilde{m}_{\alpha}^2\tilde{\Lambda} Y(\tilde{\Lambda} r)-\tilde{m}_{\alpha}^3\frac{\cos(\tilde{m}_{\alpha}r)}{\tilde{m}_{\alpha}r}\nonumber\\
&+&(\Lambda^2-m_{\alpha}^2)\tilde{\Lambda}\frac{e^{-\tilde{\Lambda} r}}{2}~]\nonumber\\
&+&\frac{1}{3}S_{12}[\tilde{m}_{\alpha}^3 Z'(\tilde{m}_{\alpha}r)+ \tilde{\Lambda}^3 Z(\tilde{\Lambda} r)\nonumber\\
&+&(\Lambda^2-m_{\alpha}^2)(1+\tilde{\Lambda} r)\frac{\tilde{\Lambda}}{2}Y(\tilde{\Lambda} r)~]\nonumber\\
&=&(\vec{\sigma_1}\cdot\vec{\sigma_2})\mathcal{F'}_{3u1\alpha} +
S_{12}\mathcal{F'}_{3u2\alpha}
\end{eqnarray}

\begin{eqnarray}
\mathcal{F}_{4t\alpha}
&=&\mathcal{F}\{{(\frac{\Lambda^2-m_{\alpha}^2}{{\Lambda}^2+\vec{q}^2})
\frac{\vec{k}^2}{\vec{q}^2+m_{\alpha}^2}}\}\nonumber\\
&=&\frac{m_{\alpha}^3}{4}Y(m_{\alpha}r)-\frac{\Lambda^3}{4}Y(\Lambda r)\nonumber\\
&-&\frac{\Lambda^2-m_{\alpha}^2}{4}(\frac{\Lambda r}{2}-1)\frac{e^{-\Lambda r}}{r}\nonumber\\
&-&\frac{1}{2}\{\nabla^2,m_{\alpha} Y(m_{\alpha}r)-\Lambda Y(\Lambda r)-\frac{\Lambda^2-m_{\alpha}^2}{2}\frac{e^{-\Lambda r}}{\Lambda}\}\nonumber\\
&=&\mathcal{F}_{4t1\alpha}+\{-\frac{1}{2}\nabla^2,\mathcal{F}_{4t2\alpha}\}
\end{eqnarray}

\begin{eqnarray}
\mathcal{F}_{4u\alpha}
&=&\mathcal{F}\{{(\frac{\Lambda^2-\tilde{m}_{\alpha}^2}{{\tilde{\Lambda}}^2+\vec{q}^2})
\frac{\vec{k}^2}{\vec{q}^2+\tilde{m}_{\alpha}^2}}\}\nonumber\\
&=&\frac{\tilde{m}_{\alpha}^3}{4}Y(\tilde{m}_{\alpha}r)-\frac{\tilde{\Lambda}^3}{4}Y(\tilde{\Lambda} r)\nonumber\\
&-&\frac{\Lambda^2-m_{\alpha}^2}{4}(\frac{\tilde{\Lambda} r}{2}-1)\frac{e^{-\tilde{\Lambda} r}}{r}\nonumber\\
&-&\frac{1}{2}\{\nabla^2,\tilde{m}_{\alpha}Y(\tilde{m}_{\alpha}r)-\tilde{\Lambda}
Y(\tilde{\Lambda} r)-\frac{\Lambda^2-m_{\alpha}^2}{2}
\frac{e^{-\tilde{\Lambda} r}}{\tilde{\Lambda}}\}\nonumber\\
&=&\mathcal{F}_{4u1\alpha}+\{-\frac{1}{2}\nabla^2,\mathcal{F}_{4u2\alpha}\}
\end{eqnarray}

\begin{eqnarray}
\mathcal{F}_{5t\alpha}
&=&\mathcal{F}\{{i(\frac{\Lambda^2-m_{\alpha}^2}{{\Lambda}^2+\vec{q}^2})
\frac{\vec{S}\cdot(\vec{q}\times\vec{k})}{\vec{q}^2+m_{\alpha}^2}}\}\nonumber\\
&=&\vec{S}\cdot\vec{L}[-m_{\alpha}^3 Z_{1}(m_{\alpha}r)+\Lambda^3 Z_{1}(\Lambda r)\nonumber\\
&+&(\Lambda^2-m_{\alpha}^2)\frac{e^{-\Lambda r}}{2r}]\nonumber\\
&=&\vec{S}\cdot\vec{L}\mathcal{F}_{5t0\alpha}
\end{eqnarray}

\begin{eqnarray}
\mathcal{F}_{5u\alpha}
&=&\mathcal{F}\{{i(\frac{\Lambda^2-m_{\alpha}^2}{{\tilde{\Lambda}}^2+\vec{q}^2})
\frac{\vec{S}\cdot(\vec{q}\times\vec{k})}{\vec{q}^2+\tilde{m}_{\alpha}^2}}\}\nonumber\\
&=&\vec{S}\cdot\vec{L}[-\tilde{m}_{\alpha}^3 Z_{1}(\tilde{m}_{\alpha}r)+\tilde{\Lambda}^3 Z_{1}(\tilde{\Lambda} r)\nonumber\\
&+&(\Lambda^2-m_{\alpha}^2)\frac{e^{-\tilde{\Lambda} r}}{2r}]\nonumber\\
&=&\vec{S}\cdot\vec{L}\mathcal{F}_{5u0\alpha}
\end{eqnarray}

\begin{eqnarray}
\mathcal{F'}_{5u\alpha}
&=&\mathcal{F}\{{i(\frac{\Lambda^2-m_{\alpha}^2}{{\tilde{\Lambda}}^2+\vec{q}^2})
\frac{\vec{S}\cdot(\vec{q}\times\vec{k})}{\vec{q}^2+\tilde{m_{\alpha}}^2}}\}\nonumber\\
&=&\vec{S}\cdot\vec{L}[-\tilde{m}_{\alpha}^3 Z'_{1}(\tilde{m}_{\alpha}r)+\tilde{\Lambda}^3 Z_{1}(\tilde{\Lambda} r)\nonumber\\
&+&(\Lambda^2-m_{\alpha}^2)\frac{e^{-\tilde{\Lambda} r}}{2r}]\nonumber\\
&=&\vec{S}\cdot\vec{L}\mathcal{F'}_{5u0\alpha}
\end{eqnarray}

\begin{eqnarray}
\mathcal{F}_{6u\alpha}&=&
\mathcal{F}\{(\frac{\Lambda^2-m_{\alpha}^2}{{\tilde{\Lambda}}^2+\vec{q}^2})
\frac{(\vec{\sigma_1}\cdot\vec{k})(\vec{\sigma_2}\cdot\vec{k})}{\vec{p}^2+\tilde{m}_{\alpha}^2}\}\nonumber\\
&=&-\frac{\vec{\sigma_1}\cdot\vec{\sigma_2}}{4}[~\tilde{m}_{\alpha}^3Y(\tilde{m}_{\alpha}r)-(\tilde{\Lambda})^3Y(\tilde{\Lambda}r)\nonumber\\
&-&(\Lambda^2-m_{\alpha}^2)\tilde{\Lambda}\frac{e^{-\tilde{\Lambda} r}}{2}~]\nonumber\\
&+&\frac{1}{3}(S_{12}+\vec{\sigma_1}\cdot\vec{\sigma_2})[~(1+\frac{3}{\tilde{m}_{\alpha}r})\tilde{m}_{\alpha}^2 Y(\tilde{\Lambda} r)\nonumber\\
&-&(1+\frac{3}{\tilde{\Lambda}r})(\tilde{\Lambda})^2 Y(\tilde{\Lambda} r)\nonumber\\
&-&(\Lambda^2-m_{\alpha}^2)(\tilde{\Lambda}+\frac{2}{r})\frac{e^{-\tilde{\Lambda} r}}{2\tilde{\Lambda}}~]\nabla\nonumber\\
&-&\frac{1}{3}(S_{12}+\vec{\sigma_1}\cdot\vec{\sigma_2})[\tilde{m}_{\alpha} Y(\tilde{m}_{\alpha}r)- \tilde{\Lambda}Y(\tilde{\Lambda} r)\nonumber\\
&-&(\Lambda^2-m_{\alpha}^2)\frac{e^{-\tilde{\Lambda} r}}{2\tilde{\Lambda}}~]\nabla^2\nonumber\\
&=&-\frac{\vec{\sigma_1}\cdot\vec{\sigma_2}}{4}\mathcal{F}_{6u1\alpha}
+\frac{1}{3}(S_{12}+\vec{\sigma_1}\cdot\vec{\sigma_2})\mathcal{F}_{6u2\alpha}\nonumber\\
&-&\frac{1}{3}(S_{12}+\vec{\sigma_1}\cdot\vec{\sigma_2})\mathcal{F}_{6u3\alpha}
\end{eqnarray}

\begin{eqnarray}
\mathcal{F'}_{6u\alpha}&=&
\mathcal{F}\{(\frac{\Lambda^2-m_{\alpha}^2}{{\tilde{\Lambda}}^2+\vec{q}^2})
\frac{(\vec{\sigma_1}\cdot\vec{k})(\vec{\sigma_2}\cdot\vec{k})}{\vec{p}^2-\tilde{m}_{\alpha}^2}\}\nonumber\\
&=&-\frac{\vec{\sigma_1}\cdot\vec{\sigma_2}}{4}[~\tilde{m}_{\alpha}^3 \frac{\cos(M_{\alpha}r)}{\tilde{m}_{\alpha}r}-(\tilde{\Lambda})^3Y(\tilde{\Lambda}r)\nonumber\\
&-&(\Lambda^2-m_{\alpha}^2)\tilde{\Lambda}\frac{e^{-\tilde{\Lambda} r}}{2}~]\nonumber\\
&+&\frac{1}{3}(S_{12}+\vec{\sigma_1}\cdot\vec{\sigma_2})[~(\frac{\sin(\tilde{m}_{\alpha}r)}{\tilde{m}_{\alpha}r}+\frac{3}{\tilde{m}_{\alpha}r}\frac{\cos(\tilde{m}_{\alpha}r)}
{\tilde{m}_{\alpha}r})\tilde{m}_{\alpha}^2\nonumber\\
&-&(1+\frac{3}{\tilde{\Lambda}r})(\tilde{\Lambda})^2
Y(\tilde{\Lambda} r)
-(\Lambda^2-m_{\alpha}^2)(\tilde{\Lambda}+\frac{2}{r})\frac{e^{-\tilde{\Lambda} r}}{2\tilde{\Lambda}}~]\nabla\nonumber\\
&-&\frac{1}{3}(S_{12}+\vec{\sigma_1}\cdot\vec{\sigma_2})[\tilde{m}_{\alpha} \frac{\cos(\tilde{m}_{\alpha}r)}{\tilde{m}_{\alpha}r}- \tilde{\Lambda}Y(\tilde{\Lambda} r)\nonumber\\
&-&(\Lambda^2-m_{\alpha}^2)\frac{e^{-\tilde{\Lambda} r}}{2\tilde{\Lambda}}~]\nabla^2\nonumber\\
&=&-\frac{\vec{\sigma_1}\cdot\vec{\sigma_2}}{4}\mathcal{F'}_{6u1\alpha}
+\frac{1}{3}(S_{12}+\vec{\sigma_1}\cdot\vec{\sigma_2})\mathcal{F'}_{6u2\alpha}\nonumber\\
&-&\frac{1}{3}(S_{12}+\vec{\sigma_1}\cdot\vec{\sigma_2})\mathcal{F'}_{6u3\alpha}
\end{eqnarray}

\end{document}